\documentclass[aps,prl,twocolumn,superscriptaddress,appendix,showpacs,floatfix,nobibnotes]{revtex4}
\usepackage{epsfig}
\usepackage{epstopdf}

\usepackage{graphicx}
\usepackage{longtable}
\usepackage{CJK}
\usepackage{color}
\usepackage{mathptmx, courier, pifont}
\usepackage[scaled=0.92]{helvet}
\usepackage[T1]{fontenc}
\usepackage{textcomp}

\begin{document}

\begin{CJK*}{GB}{}

\title{B(E2) anomaly and triaxial deformation in the interacting boson model}

\author{Yu Zhang }\email{dlzhangyu_physics@163.com}
\affiliation{Department of Physics, Liaoning Normal University,
Dalian 116029, P. R. China}

\author{Wei Teng}
\affiliation{Department of Physics, Liaoning Normal University,
Dalian 116029, P. R. China}

\begin{abstract}
The influences of triaxial dynamics on the anomalous $E2$ transitional behaviors in the interacting boson model (IBM) have been comprehensively examined without symmetry restrictions.
Specifically, different triaxial schemes in the IBM have been analyzed using a general procedure for deriving triaxial rotor modes based on two distinct approaches. The results indicate that the $B(E2)$ anomaly feature, characterized by $B_{4/2}=B(E2;4_1^+\rightarrow2_1^+)/B(E2;2_1^+\rightarrow0_1^+)<1.0$, can occur in all triaxial schemes due to strong band-mixing effects. This finding is further tested in describing the available data for $^{172}$Pt,~$^{168,170}$Os and $^{166}$W using the model constructed upon the extended consistent-$Q$ formula, providing a simple theoretical explanation of the depressed $B_{4/2}$ values observed in the experiments.
\end{abstract}

\maketitle

\end{CJK*}

\begin{center}
\vskip.2cm\textbf{I. Introduction}
\end{center}\vskip.2cm

The emergence of collective features is one of the most important and striking characteristics of complex nuclear many-body systems.
The associated collective modes can be well illustrated within a Bohr-Mottelson picture of the dynamics using geometric language~\cite{Bohrbook},
which have been extended to encompass spherical vibrator, axially-deformed rotor, triaxially-deformed rotor and $\gamma$-unstable rotor. Generally, these collective modes
can be demonstrated equivalently well within the framework of the interacting boson model (IBM) using group or algebraic language~\cite{Iachellobook},
including U(5) (spherical vibrator), SU(3) (axially-deformed rotor) and O(6) ($\gamma$-unstable rotor). In experiments,
different collective modes may be readily discerned through observing low-lying spectroscopic observables related to yrast states, such as the typical ratios,
$R_{4/2}\equiv E(4_1^+)/E(2_1^+)$ and $B_{4/2}\equiv B(E2;4_1^+\rightarrow2_1^+)/B(E2;2_1^+\rightarrow0_1^+)$. These are characterized by $R_{4/2}\equiv E(4_1^+)/E(2_1^+)\approx2.0$ for spherical vibrator, $R_{4/2}\approx3.33$ for axially-deformed rotor and $R_{4/2}\approx2.5$ for $\gamma$-unstable rotor. Although the different modes are not easy to distinguish from each other by using the $B(E2)$ ratio, they all show a common feature of $B_{4/2}\equiv B(E2;4_1^+\rightarrow2_1^+)/B(E2;2_1^+\rightarrow0_1^+)>1.0$, along with $R_{4/2}\geq2$. For a long time this has been regarded as a robust rule governing nuclear collective modes due to its consistency with various theoretical calculations and extensive experimental data on collective nuclei.
However, this rule has been broken by observing the results from the recent measurements on some neutron-deficient nuclei~\cite{Grahn2016,Saygi2017,Cederwall2018,Goasduff2019,Zhang2021} near $N_\mathrm{n}=90$ and the proton dripline, which suggest an anomalous collective motion characterized by $R_{4/2}>2.0$ and $B_{4/2}\ll1.0$. The unusually $B_{4/2}<1.0$ feature was actually found even earlier to exist in other collective nuclei but to a less extent~\cite{Cakirli2004}. Clearly, such an anomalous phenomenon, called $B(E2)$ anomaly, cannot be understood from the familiar conventional collective modes nor be produced by large-scale shell model or beyond mean-field approaches, therefore leaving a big challenge to their theoretical explanations~\cite{Grahn2016}.

Recent progress on this issue has been made~\cite{Zhang2022} within the IBM framework by employing the SU(3) image of the triaxial rotor~\cite{Zhang2014,Smirnov2000,Leschber1987,Castanos1988}, by which the $B(E2)$ anomaly characterized by $B_{4/2}<1.0$ is attributed as a result of the finite-$N$ triaxial rotor modes~\cite{Zhang2014}. This method, which introduces high-order SU(3) symmetry-conserving terms, may extend the traditional descriptions of axially-deformed (prolate or oblate) rotor, based on two-body SU(3) interactions, to triaxially-deformed rotor cases~\cite{Zhang2014,Smirnov2000}. However, another work~\cite{Wang2023} pointed out that the Hamiltonian established in the O(6) limit, analogous to that adopted in the SU(3) image~\cite{Zhang2022}, fails to produce $B(E2)$ anomaly. This raises an question: can the $B(E2)$ anomaly feature manifests in an IBM scenario different from the SU(3) scheme or a specific type of Hamiltonian~\cite{Zhang2022}. In other words, to what extent is the $B(E2)$ anomaly feature dictated by triaxial deformation a model-independent result. On the other side, microscopic mean-field calculations suggest that axially asymmetric deformations can occur in nuclear systems~\cite{Xiang2018}, and it has even been pointed out that some degree of triaxiality is present in nearly all regions of the nuclear chart according to the recent analysis based on the shell model approaches~\cite{Bonatsos2017I,Bonatsos2017II,Tsunoda2021,Otsuka2023,Rouoof2024}. In contrast, a mean-field $\gamma$ deformation has no way to be generated by the frequently adopted IBM Hamiltonian up to two-body terms~\cite{VC1981}. So, establishing triaxial rotor modes within the IBM framework associated with a certain of intrinsic deformation remains an interesting objective. The previous approach used to explain the $B(E2)$ anomaly~\cite{Zhang2022} was just constructed from the IBM realization of triaxial rotor modes~\cite{Zhang2014,Smirnov2000}, which was achieved based on the SU(3) image of the rotor model~\cite{Leschber1987} and guided by the relationship between the $\gamma$ deformation and SU(3) irreducible representations (IRREPs)~\cite{Castanos1988}. Nevertheless, there is more than one way to produce triaxial deformation in the IBM, such as introducing the cubic interaction $(d^\dag\times d^\dag\times d^\dag)^{(3)}\cdot(\tilde{d}\times\tilde{d}\times\tilde{d})^{(3)}$~\cite{VC1981,Heyde1984} or adopting the extended consistent-$Q$ formula~\cite{Fortunato2011}. These approaches were ever widely discussed to identify the influence of triaxiality in the IBM descriptions of nuclear properties~\cite{VC1981,Heyde1984,Fortunato2011,Sorgunlu2008,Casten1985,Ramos2000I,Ramos2000II,Wang2023} and will be revisited here to uncover the correlation between $B(E2)$ anomaly and triaxial deformation.

The main purpose of this work is to provide a new procedure of constructing triaxial modes in the IBM, by which different triaxial schemes allowing $B_{4/2}<1.0$ will be examined to clarify the role of triaxiality in generating $B(E2)$ anomaly.

\begin{center}
\vskip.2cm\textbf{II. Triaxial rotor Hamiltonian in the IBM}
\end{center}\vskip.2cm

Hamiltonian and physical operators in the IBM~\cite{Iachellobook} are constructed from two kinds of boson operators, namely the $s$ boson with $J^\pi=0^+$
and the $d$ boson with $J^\pi=2^+$.
Among them, the angular momentum operator and quadrupole moment operator are defined by
\begin{eqnarray}\label{QLI}
&&\hat{L}_u=\sqrt{10}(d^\dag\times\tilde{d})_u^{(1)}\, ,\\ \label{QLII}
&&\hat{Q}_u=(d^\dag\times\tilde{s}+s^\dag\times\tilde{d})_u^{(2)}+\chi(d^\dag\times\tilde{d})_u^{(2)}\,
\end{eqnarray}
with the parameter $\chi\in[-\frac{\sqrt{7}}{2},~\frac{\sqrt{7}}{2}]$.
To build triaxial modes in the IBM, the Hamiltonian is proposed to include two parts,
\begin{eqnarray}\label{Tri}
\hat{H}_{\mathrm{Tri}}=\hat{H}_\mathrm{S}+\hat{H}_\mathrm{D}\, ,
\end{eqnarray}
which is achieved by analogy with the form adopted by the SU(3) scheme~\cite{Zhang2022}.
The static part $\hat{H}_\mathrm{S}$ is designed to yield a triaxial potential at the mean-field level, while the dynamic part $\hat{H}_\mathrm{D}$
serves to generate the associated band-mixing effects. Specifically, $\hat{H}_\mathrm{D}$ can be
formulated from a scalar polynomial in $\hat{L}$ and $\hat{Q}$ and given by
\begin{eqnarray}\label{HD}
\hat{H}_\mathrm{D}=a\hat{L}^2+b(\hat{L}\times \hat{Q}\times \hat{L})^{(0)}+c(\hat{L}\times \hat{Q})^{(1)}\cdot(\hat{L}\times \hat{Q})^{(1)}\, ,
\end{eqnarray}
where $a,~b,~c$ denote the parameters related to the specified triaxial deformation as discussed below.
It is noteworthy that the terms involved in (\ref{HD}) have been extensively addressed in the IBM description of rotational dynamics~\cite{Zhang2014,Smirnov2000,Berghe1985,Vanthournout1988,Vanthournout1990,Teng2024}. For instance, it has been demonstrated that the degeneracies in members of the $\beta$ and $\gamma$ bands in the SU(3) limit of the IBM could be lifted by incorporating these terms into the Hamiltonian~\cite{Vanthournout1990}.

\begin{center}
\vskip.2cm\textbf{(A). Relation to the rotor model}
\end{center}\vskip.2cm

\begin{figure}
\begin{center}
\includegraphics[scale=0.35]{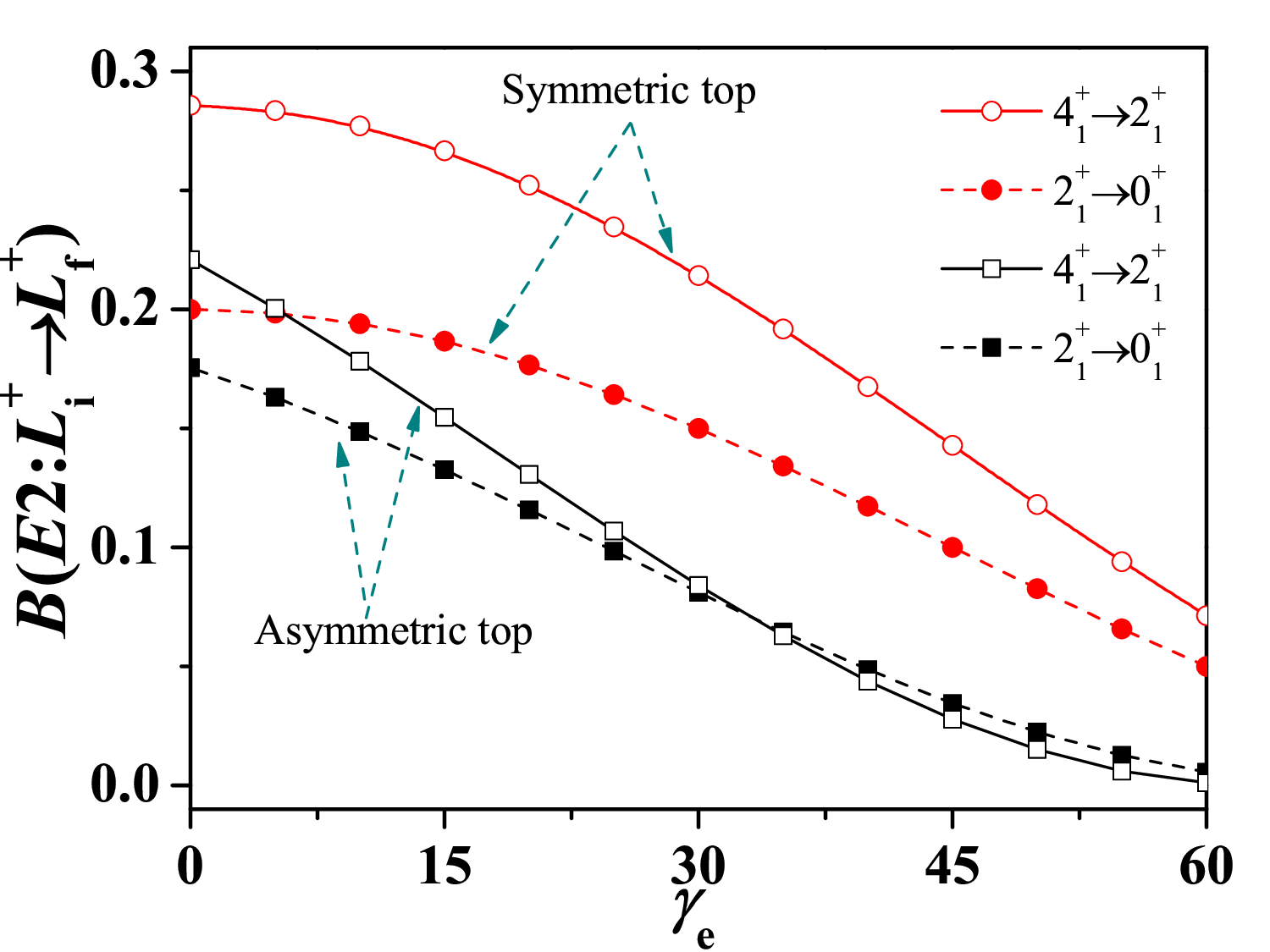}
\caption{The values of $B(E2;2_1^+\rightarrow 0_1^+)$ and $B(E2;4_1^+\rightarrow 2_1^+)$ (in any units) derived from the rotor model described by (\ref{Hr}) are presented as a function of the values (in degree) of $\gamma_\mathrm{e}$, which is incorporated into the electronic quadrupole operator $T^{E2}=e\beta[\mathrm{cos}(\gamma_\mathrm{e})D_{u,0}+\frac{1}{\sqrt{2}}\mathrm{sin}(\gamma_\mathrm{e})(D_{u,2}+D_{u,-2})]$, where the effective charge $e$ has been fixed by $e\beta=1.0$ for convenience in computing $B(E2)$ transitions. In these calculations, two sets of inertia parameters have been selected: $\Gamma_1:\Gamma_2:\Gamma_3=1:1:4$ to represent symmetric tops, and $\Gamma_1:\Gamma_2:\Gamma_3=3:1:4$ for asymmetric tops. Similarly to the discussion in \cite{Leschber1987}, no assumptions are made regarding a relationship between the initial parameters $\Gamma_\alpha$ and the deformation parameter $\gamma_\mathrm{e}$. }\label{F0}
\end{center}
\end{figure}

To illustrate the role of $\hat{H}_\mathrm{D}$ defined in (\ref{HD}), we will demonstrate how the Hamiltonian form can be related to a triaxial rotor.
The Hamiltonian of the quantum rotor can generally be written as
\begin{equation}\label{Hr}
H_{\mathrm{rot}}=\Gamma_1L_1^2+\Gamma_2L_2^2+\Gamma_3L_3^2\, ,
\end{equation}
where $L_\alpha$ with $\alpha=1,~2,~3$ (or $x,~y,~z$) represents the projection of the angular momentum onto the
$\alpha$-th body-fixed principal axis and $\Gamma_\alpha$ denotes the
corresponding inertia parameter. By taking $\Gamma_2\leq\Gamma_1\leq\Gamma_3$, which aligns with the order generated by the rigid moment of inertia~\cite{Zhang2014}, it is convenient to
classify different rotational tops by introducing an asymmetry parameter related to the inertia ellipsoid, as proposed in \cite{Leschber1987}. The parameter is defined as $\bar{k}=(2\Gamma_1-\Gamma_3-\Gamma_2)/(\Gamma_3-\Gamma_2)$ with range $-1\leq \bar{k}\leq1$. Consequently, the $\bar{k}=-1$ limit corresponds to a prolate top with $\Gamma_1=\Gamma_2<\Gamma_3$, while $\bar{k}=1$ characterizes an oblate top with $\Gamma_1=\Gamma_3>\Gamma_2$, and
$\bar{k}=0$ indicates the most asymmetric case with $\Gamma_1=(\Gamma_2+\Gamma_3)/2$.
For an asymmetric top (triaxial rotor) with $-1<\bar{k}<1$, numerical calculations reveal that the associated strong band-mixing effects can significantly reduce the $B_{4/2}$ ratio, potentially down to values less than 1.0. An example illustrating these results is presented in the Fig.~\ref{F0}, where it can be observed that the cases yielding $B_{4/2}<1.0$ arise from asymmetric tops when $\gamma_\mathrm{e}>30^\circ$. In contrast, symmetric tops characterized by $\bar{k}=\pm1$ constantly produce $B_{4/2}>1.0$ within the range of $0^\circ\leq\gamma_\mathrm{e}\leq60^\circ$. Here, $\gamma_\mathrm{e}$ pertains to the parameter involved in the electronic quadrupole moment (see Fig.~\ref{F0}), and no assumptions are made regarding relationships between  $\gamma_\mathrm{e}$ and initial parameters $\Gamma_\alpha$~\cite{Leschber1987}, which are taken as independent parameters for calculating $B(E2)$ transitions. Although the numerical results depicted in Fig.~\ref{F0} provide only a schematic representation that may not correspond directly to actual situations, the resulting model features offer valuable insights for explaining $B(E2)$ anomaly from the perspective of triaxial rotor modes. However, the rigid triaxial deformation dictated by the rotor model is not anticipated to genuinely occur in realistic systems, where low-lying dynamics typically involve multiple types of collective modes. Furthermore, the finite nucleon number effects, which is considered an important factor for accurately describing various collective phenomena~\cite{Casten1983,Casten2016}, cannot be captured by the rotor model Hamiltonian. In contrast, the IBM offers a suitable framework to address these limitations. In what follows, we will demonstrate how a triaxial rotor mode can be incorporated in an apparent way using the Hamiltonian (\ref{HD}), based on the strategy previously proposed in \cite{Leschber1987,Castanos1988}. This method bas been successfully applied in developing the pseudo-SU(3) shell model for heavy nuclei~\cite{Draayer1989,Draayer1983} and further extended to the SU(3) IBM model~\cite{Zhang2014,Zhang2022}.

According to the analysis given in \cite{Leschber1987}, the quantum rotor Hamiltonian (\ref{Hr}) can be rewritten as
\begin{equation}\label{HLQ}
H_{\mathrm{rot}}=a^\prime L^2+b^\prime X_3^\mathrm{c}+c^\prime X_4^\mathrm{c}\, ,
\end{equation}
where $a^\prime, b^\prime, c^\prime$ represent the coefficients to be determined
and the three frame-independent scalars are constructed by
\begin{eqnarray}\label{LXX1}
&&L^2=L_1^2+L_2^2+L_3^2\, ,\\ \label{LXX2}
&&X_3^\mathrm{c}=\sum_{\alpha\beta}L_\alpha Q_{\alpha\beta}^\mathrm{c} L_\beta=\lambda_1L_1^2+\lambda_2L_2^2+\lambda_3L_3^2\, ,\\ \label{LXX3}
&&X_4^\mathrm{c}=\sum_{\alpha\beta\gamma}L_\alpha Q_{\alpha\beta}^\mathrm{c}Q_{\beta\gamma}^\mathrm{c} L_\gamma=\lambda_1^2L_1^2+\lambda_2^2L_2^2+\lambda_3^2L_3^2\, .
\end{eqnarray}
Here, $L_\alpha$ and $Q_{\alpha\beta}^\mathrm{c}$ with ($\alpha,\beta,\gamma$) being cyclic permutation of ($1,2,3$) or ($x,y,z$) represent the cartesian form of the angular momentum operators and the (mass) quadrupole moment operators, which are defined as follows,
\begin{eqnarray}\label{LQdefintion}
&&L_{\alpha}=\int\rho(r)(x_\beta v_\gamma-x_\gamma v_\beta)d^3r,\\
&&Q_{\alpha\beta}^\mathrm{c}=\int\rho(r)(3x_\alpha x_\beta-r^2\delta_{\alpha\beta})d^3r\, ,
\end{eqnarray}
where $v_\alpha=-i\frac{\partial}{\partial x_\alpha}$ and $\rho(r)$ represents the nuclear density.
The expectation values of the quadruple moment in the principal-axes system can be expressed as $\langle Q_{\alpha\beta}^\mathrm{c}\rangle=\lambda_\alpha\delta_{\alpha\beta}$,
as demonstrated in the last expression of each equation given
in (\ref{LXX1})-(\ref{LXX3}). Accordingly, one can derive that
\begin{equation}\label{Ia}
L_{\alpha}^2=[(\lambda_1\lambda_2\lambda_3)L^2+\lambda_\alpha^2X_3^\mathrm{c}+\lambda_\alpha
X_4^\mathrm{c}]/(2\lambda_\alpha^3+\lambda_1\lambda_2\lambda_3)\, .
\end{equation}
Substituting this result into (\ref{Hr}), one can get the exact relation of the coefficients,
\begin{eqnarray}\label{abc}\nonumber
&&a^\prime=\sum_\alpha a_\alpha
\Gamma_\alpha,~~~~a_\alpha=\lambda_1\lambda_2\lambda_3/D_\alpha\, ,\\
&&b^\prime=\sum_\alpha b_\alpha
\Gamma_\alpha,~~~~b_\alpha=\lambda_\alpha^2/D_\alpha\, ,\\ \nonumber
&&c^\prime=\sum_\alpha c_\alpha
\Gamma_\alpha,~~~~c_\alpha=\lambda_\alpha/D_\alpha\, ,
\end{eqnarray}
and
\begin{equation}
D_\alpha=2\lambda_\alpha^3+\lambda_1\lambda_2\lambda_3\, .
\end{equation}
With the spherical tensor formulas~\cite{Teng2024},
\begin{eqnarray}\label{LQspherical1}
&&L_0=L_z,\\ \label{LQspherical2}
&&L_{\pm1}=\mp\frac{1}{\sqrt{2}}(L_x\pm iL_y),\\ \label{LQspherical3}
&&Q_0^\mathrm{c}=3Q_{zz}^\mathrm{c},\\ \label{LQspherical4}
&&Q_{\pm1}^\mathrm{c}=\mp\sqrt{6}(Q_{xz}^\mathrm{c}\pm iQ_{yz}^\mathrm{c}),\\ \label{LQspherical5}
&&Q_{\pm2}^\mathrm{c}=\sqrt{\frac{3}{2}}(Q_{xx}^\mathrm{c}-Q_{yy}^c\pm 2iQ_{xy}^\mathrm{c})\, ,
\end{eqnarray}
the frame-independent scalars presented in (\ref{LXX1})-(\ref{LXX3}) can be rewritten as
\begin{eqnarray}\label{a}
&&L^2=\sqrt{5}(L\times L)^{(0)}\, ,\\ \label{b}
&&X_3^\mathrm{c}=\frac{\sqrt{30}}{6}(L\times Q^\mathrm{c}\times L)^{(0)}\, ,\\ \label{c}
&&X_4^\mathrm{c}=\frac{5}{18}(L\times Q^\mathrm{c})^{(1)}\cdot(L\times Q^\mathrm{c})^{(1)}\, .
\end{eqnarray}
This is referred to as the algebraic realization of triaxial rotor~\cite{Leschber1987}. Although the Hamiltonian (\ref{Hr}) and (\ref{HLQ}) may yield identical eigenvalues and eigenvectors based on the coefficients relations specifies in (\ref{abc}), the latter formulation is more readily applicable within other frameworks, such as the IBM, since it comprises only the scalar functions of angular momentum and quadrupole moment operators.

One can prove~\cite{Ui1970} that
the angular momentum operators and quadrupole moment operators defined in (\ref{LQspherical1})-(\ref{LQspherical5}) may generate the semidirect sum  Lie algebra $\mathrm{t_5}\oplus_\mathrm{s} \mathrm{so(3)}$ and satisfy the commutation relations
\begin{eqnarray}\label{LL}
&&[L_u,L_v]=-\sqrt{2}\langle1u,1v|1u+v\rangle L_{u+v}\, ,\\ \label{LQ}
&&[L_u,Q_v^\mathrm{c}]=-\sqrt{6}\langle1u,2v|2u+v\rangle Q_{u+v}^\mathrm{c}\, ,\\
&&[Q_u^\mathrm{c},Q_v^\mathrm{c}]=0\, .
\end{eqnarray}
Meanwhile, one can construct the invariant operators~\cite{Castanos1988} as
\begin{eqnarray}\label{Qt1}
&&C_2=\frac{\sqrt{5}}{6}(Q^\mathrm{c}\times Q^\mathrm{c})^{(0)},\\ \label{Qt2}
&&C_3=-\frac{1}{36}\sqrt{\frac{35}{2}}(Q^\mathrm{c}\times Q^\mathrm{c}\times Q^\mathrm{c})^{(0)}\, ,
\end{eqnarray}
which satisfy $[L_u,~C_k]=[Q_u^\mathrm{c},~C_k]=0$ with $k=2,3$.
In the body-fixed, principal-axes system,
the eigenvalues are given by
\begin{eqnarray}\label{lambda1}
&&\langle C_2\rangle=\lambda_1^2+\lambda_2^2+\lambda_3^2,\\ \label{lambda2}
&&\langle C_3\rangle=\lambda_1^3+\lambda_2^3+\lambda_3^3=3(\lambda_1\lambda_2\lambda_3)\, .
\end{eqnarray}
Here, $\lambda_\alpha$ with $\alpha=1,2,3$ are just the expectation values of the quadrupole matrix described above. Given that $Q^\mathrm{c}$ is a traceless tensor with $\lambda_1+\lambda_2+\lambda_3=0$, it can be further established that $\lambda_\alpha$ satisfies the cubic equation~\cite{Rosensteel1977}
\begin{equation}\label{cubic}
\lambda_\alpha^3-\frac{1}{2}\langle C_2\rangle\lambda_\alpha-\frac{1}{3}\langle C_3\rangle=0\, .
\end{equation}
The solutions can be uniformly expressed as~\cite{Castanos1988}
\begin{eqnarray}\label{solution}
&&\lambda_\alpha=2\sqrt{\frac{\langle C_2\rangle}{6}}\mathrm{cos}\Big(\frac{\theta}{3}-\frac{2\alpha\pi}{3}\Big),\\
&&\theta=\mathrm{tan}^{-1}\Big[\frac{\langle C_2\rangle^3}{6\langle C_3\rangle^2}-1\Big]^{1/2}\, ,
\end{eqnarray}
where $\langle C_2\rangle^3\geq6\langle C_3\rangle^2$ is required to ensure the existence of real solutions.
Clearly, if a method can be established to determine $\langle C_2\rangle$ and $\langle C_3\rangle$, then the values of $\lambda_\alpha$ can be directly
obtained from (\ref{solution}).

In the IBM calculations, the angular momentum operator $L$ and quadrupole operator $Q^\mathrm{c}$ are approximately replaced by those defined in (\ref{QLI})-(\ref{QLII}), with an additional scaling factor of $2\sqrt{2}$ included in $\hat{Q}$ to ensure the consistent commutation relations between $\hat{L}$ and $\hat{Q}$, as specified in (\ref{LQ}). Consequently, it becomes evident that the Hamiltonian presented in (\ref{HD}) represents merely an IBM analogue of the rotor Hamiltonian (\ref{HLQ}), with the parameter relations specified as $a=a^\prime,~b=\frac{2\sqrt{15}}{3}b^\prime$ and $c=\frac{20}{9}c^\prime$. Therefore, introducing $\hat{H}_\mathrm{D}$ here serves primarily to establish the triaxial rotor modes in the IBM rather than bring high-order improvements of rotational dynamics. The parameters associated with $\hat{H}_\mathrm{D}$ corresponding to a rotor mode describe by (\ref{Hr}) with
the specific $\Gamma_\alpha$ will be fully constrained by the mapping formulas given in (\ref{abc}), once $\langle C_2\rangle$ and $\langle C_3\rangle$ are evaluated. As demonstrated below, two distinct approaches are proposed for deriving $\hat{H}_\mathrm{D}$ within the IBM.

\begin{center}
\vskip.2cm\textbf{(B). Two approaches for deriving the rotor modes}
\end{center}\vskip.2cm

The first approach is to approximately evaluate $\langle C_2\rangle$ and $\langle C_3\rangle$ in the classical (large-$N$) limit of the IBM.
To achieve this, it is essential to identify the intrinsic deformation of a given system using the coherent state method~\cite{VC1981}.
Specifically, one can determine the classical limit of an IBM Hamiltonian through the coherent state defined by~\cite{Iachellobook}
\begin{eqnarray}\label{coherent}
|\beta, \gamma, N\rangle=G[s^\dag + \beta \mathrm{cos} \gamma~
d_0^\dag\ + \frac{1}{\sqrt{2}} \beta \mathrm{sin} \gamma (d_2^\dag +
d_{ - 2}^\dag)]^N |0\rangle\,
\end{eqnarray}
with the normalization factor $G=1/\sqrt{N!(1+\beta^2)^N}$.
The corresponding classical potential for $\hat{H}_{\mathrm{S}}$ is then defined as
\begin{equation}
V_{\,\mathrm{S}}(\beta,\gamma)=\frac{1}{N}\langle\beta, \gamma, N|\hat{H}_{\mathrm{S}}|\beta, \gamma,N\rangle|_{N\rightarrow\infty}\, ,
\end{equation}
from which we obtain the ground-state energy per boson by minimizing the potential with respect to $\beta$ and $\gamma$, giving $e_\mathrm{g}=V_{\mathrm{min}}=V_{\,\mathrm{S}}(\beta_0,\gamma_0)$.
Note that the variables $\beta$ and $\gamma$ in the IBM are related to those introduced by Bohr and Mottelson in the geometric model, denoted as $\beta_\mathrm{BM}$ and $\gamma_\mathrm{BM}$, by
$\gamma_\mathrm{BM}=\gamma$ and $\beta_\mathrm{BM}=t\beta$, where $t$ is a dimensionless factor typically with $t\sim0.16$ for rare earth nuclei~\cite{Iachellobook}.
The potential $V_{\mathrm{S}}(\beta,\gamma)$ simultaneously describes the classical limit of the full triaxial Hamiltonian $\hat{H}_{\mathrm{Tri}}$ since its dynamical part $\langle \hat{H}_\mathrm{D}\rangle/N$ in the large-$N$ limit is supposed to contribute nothing to the ground-state energy~\cite{Zhang2022}.
With the obtained optimum values ($\beta_0, \gamma_0$), one can approximately evaluate $\langle C_2\rangle$ and $\langle C_3\rangle$ at the mean-field level~\cite{Teng2024} by
substituting the $\hat{Q}$ operator defined in the IBM as given in (\ref{QLII}) into the definitions (\ref{Qt1})-(\ref{Qt2}). Consequently, the values of $\lambda_\alpha$ will be solved from Eq.~(\ref{solution}) and then applied to derive the algebraic counterpart of the rotor Hamiltonian via relation Eq.~(\ref{abc}). In short, once the static part Hamiltonian $\hat{H}_{\mathrm{S}}$ is provided along with its mean-field deformation determined through coherent methodology, the triaxial Hamiltonian in the IBM associated with a given rotor mode will be completely specified. This procedure for deriving the triaxial rotor modes can be applied universally across quadrupole systems, without being restricted to specific symmetries like the SU(3) scheme~\cite{Zhang2014,Zhang2022}. Its another advantage lies in that intrinsic deformation of a given quadrupole system can be directly incorporated when determining the rotor modes.

The second approach is to evaluate the expectation values $\langle C_2\rangle_\mathrm{g}$ and $\langle C_3\rangle_\mathrm{g}$ using the ground state wave function solved from the static Hamiltonian $\hat{H}_{\mathrm{S}}$, as the dynamic part Hamiltonian $\hat{H}_{\mathrm{D}}$ may contribute nothing to all states with $L=0$, including the ground state. In this process, the operator $\hat{Q}$ involved in the calculations will be replaced by that defined in the IBM, as specified in (\ref{QLII}). Subsequently, $\lambda_\alpha$ in (\ref{abc}) can similarly be extracted from Eq.~(\ref{solution}), allowing for the derivation of the dynamic Hamiltonian $\hat{H}_{\mathrm{D}}$ through the mapping formulas (\ref{abc}). In this case, the effective $\gamma$ deformation can be determined using a formula suggested previously~\cite{Elliott1986}, which is given as
\begin{eqnarray}\label{effgamma}
\mathrm{Cos}(3\gamma_{\mathrm{eff.}})=-\Big(\frac{7}{2\sqrt{5}}\Big)^{1/2}\frac{\langle(\hat{Q}\times\hat{Q}\times\hat{Q})^{(0)}\rangle_\mathrm{g}}{(\langle\hat{Q}\times\hat{Q})^{(0)}\rangle_\mathrm{g}^{3/2}}\, .
\end{eqnarray}
Likewise, this approach is applicable to any IBM systems to investigate the influence of triaxial rotor modes, even for those described by the consistent-$Q$ Hamiltonian (discussed later) that do not exhibit any intrinsic asymmetric deformation at the mean-field level. In such cases, triaxiality inferred from Eq.~(\ref{effgamma}) represents a dynamic result rather than one arising from intrinsic $\gamma$ deformation as discussed in the first approach. Therefore, the effective triaxial deformation defined here as $0^\circ<\gamma_{\mathrm{eff.}}<60^\circ$ based on Eq.~(\ref{effgamma}) is referred to as dynamic triaxial deformation and, to some extent, embodies the average effect of $\gamma$ deformation~\cite{Vogel1996,Castanos1984}. It has been demonstrated~\cite{Vogel1996} that this definition of effective triaxial deformation can be applied to signify triaxialilty in $\gamma$-soft systems, such as Xe and Ba nuclei with $A\sim130$. This observation can be partially understood from the fact that the predictions for $\gamma$-soft and $\gamma$-rigid potentials are nearly identical for most observables if the average value of $\gamma$ in the first case, $\gamma_{\mathrm{av}}$, is equal to the fixed value of $\gamma$ in the second, as noted in ~\cite{Zamfir1991}.
If the static Hamiltonian $\hat{H}_\mathrm{S}$ can develop a well-pronounced $\gamma$ deformation in its mean-field potential, the intrinsic $\gamma$ deformation inferred from the coherent state method will be approximately equivalent to the effective $\gamma$ deformation (or to say dynamic $\gamma$ deformation), showing $\gamma_0\approx\gamma_{\mathrm{eff.}}$, even though the former has a clearer geometric meaning. In this case, the rotor modes derived from both approaches will exhibit similar rigid-rotor dynamics.

Realistic nuclear systems at low spins are rarely $\gamma$-rigidly deformed and their low-lying structures often relate more closely to soft rotor modes or transitional cases between rotation and vibration. Even so, the proposed approaches for deriving rotor modes offer an effective means to constrain parameters in $\hat{H}_\mathrm{D}$ for weakly deformed systems. Such constraints typically ensure that a regular level pattern emerges theoretically without necessitating fine-tuning of model parameters.

\begin{center}
\vskip.2cm\textbf{III. Schemes for triaxial deformation}
\end{center}\vskip.2cm

As noted in \cite{Zhang2022}, $B(E2)$ anomaly is relatively easier to produce from a triaxially deformed system, which imposes stringent requirements for choosing Hamiltonian $\hat{H}_{\mathrm{S}}$.
In fact, achieving stable triaxial deformations in the IBM had been a active issue of research since early studies aimed at bridging between geometric models and IBM. It has been established~\cite{VC1981} that only the inclusion of higher order terms in the IBM Hamiltonian can lead to triaxial equilibrium shapes. Previously, how to build $\hat{H}_{\mathrm{S}}$ with arbitrary $\gamma$ deformation by introducing high-order SU(3) symmetry-conserving terms has been discussed in details~\cite{Smirnov2000,Zhang2014,Zhang2022}. In what follows, we will explore three additional schemes for generating intrinsic triaxial deformations in the IBM, denoted as (A), (B) and (C), respectively. A common feature among these schemes is that all exhibit a dependence on $\mathrm{cos}^2(3\gamma)$ in their classical potential functions, indicating the feasibility of triaxial deformation at the mean-field level. On the other hand, the above discussions suggest that effective triaxial deformation can be achieved via Eq.~(\ref{effgamma}), even when using an IBM Hamiltonian not directly related to intrinsic $\gamma$ deformation. An example illustrating dynamical triaxiality in such a situation will be provided in subsection (D).

\begin{center}
\vskip.2cm\textbf{(A). The O(6) scalar polynomial}
\end{center}\vskip.2cm

In the O(6) limit, it is possible to achieve triaxial deformation by introducing the O(6) symmetry-conserving scalar polynomial in the operator $\hat{Q}$ up to the sixth order~\cite{Teng2024,Thiamova2010}. Specifically, the static Hamiltonian $\hat{H}_\mathrm{S}$ in this case is constructed as
\begin{eqnarray}\label{HSA}\nonumber
&&\hat{H}_\mathrm{S_A}=A_1\frac{\hat{Q}\cdot \hat{Q}}{N}+A_2\frac{(\hat{Q}\times \hat{Q}\times \hat{Q})^{(0)}}{N(N-1)}+A_3\frac{(\hat{Q}\cdot \hat{Q})^2}{N(N-1)(N-2)}\\
&&~~~~~+A_4\frac{(\hat{Q}\times \hat{Q}\times \hat{Q})^{(0)}\cdot(\hat{Q}\times \hat{Q}\times \hat{Q})^{(0)}}{N(N-1)(N-2)(N-3)(N-4)}\, ,
\end{eqnarray}
where $A_i$ with $i=1,~2,~3,~4$ are real parameters.
If the $\hat{Q}$ operator is considered to defined in the SU(3) limit, corresponding to the definition (\ref{QLII}) with $\chi=-\sqrt{7}/2$, the first three terms in (\ref{HSA}) suffice for
generating stable triaxial deformation~\cite{Smirnov2000,Zhang2014,Zhang2022}.
However, we adopt here the O(6) definition as given in (\ref{QLII}) with $\chi=0$, whereby these initial three terms may yield either an axially deformed or $\gamma$-unstable deformed configuration in the large-$N$ limit.
Using the coherent state defined in (\ref{coherent}), the classical potential function corresponding to the Hamiltonian (\ref{HSA}) is derived as
\begin{eqnarray}\label{VSA}
V_{\,\mathrm{S_A}}(\beta,\gamma)&=&\frac{1}{N}\langle\beta, \gamma,
N|\hat{H}_{\mathrm{S_A}}|\beta, \gamma,N\rangle|_{N\rightarrow\infty}\\
\nonumber
&=&A_1\frac{4\beta^2}{(1+\beta^2)^2}-A_2\frac{16\sqrt{70}\beta^3\mathrm{cos}(3\gamma)}{70(1+\beta^2)^3}\\
\nonumber
&+&A_3\frac{16\beta^4}{(1+\beta^2)^4}+A_4\frac{16^2\beta^6\mathrm{cos}^2(3\gamma)}{70(1+\beta^2)^6}\, .
\end{eqnarray}
Clearly, the terms $\hat{Q}\cdot \hat{Q}$ and $(\hat{Q}\cdot \hat{Q})^2$ can only yield a $\gamma$-independent mean-field structure indicating $\gamma$ unstable, while $(\hat{Q}\times \hat{Q}\times \hat{Q})^{(0)}$ may describe an axially-deformed situation~\cite{Isacker1999,Rowe2005}, associated with $\gamma_0=0^\circ$ or $60^\circ$, due to its dependence on $\mathrm{cos}(3\gamma)$.
In contrast, the six-order term $(\hat{Q}\times \hat{Q}\times \hat{Q})^{(0)}\cdot(\hat{Q}\times \hat{Q}\times \hat{Q})^{(0)}$ contributes a component proportional to $\mathrm{cos}^2(3\gamma)$~\cite{Thiamova2010}, which may introduce triaxial deformation at the mean-field level~\cite{VC1981}. Once the ground-state deformation, ($\beta_0,\gamma_0$), is known, all other parameters in $\hat{H}_\mathrm{Tri}$ can be determined based on the aforementioned procedure. It should be mentioned that the fifth-order term in $\hat{Q}$ does contribute any new dependence on $\gamma$ within the classical potential and is therefore omitted here for simplicity.

\begin{center}
\vskip.2cm\textbf{(B). The cubic interaction}
\end{center}\vskip.2cm

It has been long known~\cite{VC1981} that the cubic interactional term $(d^\dag\times d^\dag\times d^\dag)^{(3)}\cdot(\tilde{d}\times\tilde{d}\times\tilde{d})^{(3)}$
can induce triaxial deformation in its classical potential function. Specifically, a stable triaxial minimum at the mean-field level
can be achieved by incorporating this term into the O(6) limit of the IBM~\cite{Heyde1984,Sorgunlu2008}. The associated dynamical effects have been demonstrated to enhance the traditional IBM descriptions of the Xe, Ba, Os, Er isotopes etc~\cite{Heyde1984,Sorgunlu2008,Casten1985,Ramos2000I,Ramos2000II}. In this context, the static Hamiltonian is formulated as
\begin{eqnarray}\label{HSB}
\hat{H}_\mathrm{S_B}=B_1\frac{\hat{Q}\cdot \hat{Q}}{N}+B_2\frac{(d^\dag\times d^\dag\times d^\dag)^{(3)}\cdot(\tilde{d}\times\tilde{d}\times\tilde{d})^{(3)}}{N(N-1)}\,
\end{eqnarray}
with the $\hat{Q}$ operator defined in accordance with the O(6) limit, similar to the scheme (A).
Consequently, the classical potential function is derived as
\begin{eqnarray}\label{VSB}
V_{\,\mathrm{S_B}}(\beta,\gamma)&=&\frac{1}{N}\langle\beta, \gamma,
N|\hat{H}_{\mathrm{S_B}}|\beta, \gamma,N\rangle|_{N\rightarrow\infty}\\
\nonumber
&=&B_1\frac{4\beta^2}{(1+\beta^2)^2}+B_2\frac{\beta^6[\mathrm{cos}^2(3\gamma)-1]}{7(1+\beta^2)^3}\, .
\end{eqnarray}
Clearly, this potential function also includes a term dependent on $\mathrm{cos}^2(3\gamma)$, indicating that triaxial deformation may be obtained in this scenario.

\begin{center}
\vskip.2cm\textbf{(C). The extended consistent-$Q$ Hamiltonian}
\end{center}\vskip.2cm

A natural extension~\cite{Fortunato2011} of the widely utilized consistent-$Q$ formulism~\cite{Warner1983} is to introduce a cubic combination of $\hat{Q}$ to the Hamiltonian.
An intriguing aspect of this extension is that it not only provides an alternative way for describing the prolate-oblate shape phase transition~\cite{Zhang2012,Wang2023II}, distinct from traditional approach~\cite{Jolie2001,Moreno1996}, but also allows triaxial deformation at the mean-field level~\cite{Fortunato2011}.
Specifically, the extended consistent-$Q$ Hamiltonian can be expressed as
\begin{eqnarray}\label{HSC}
\hat{H}_\mathrm{S_C}=C_1\hat{n}_d+C_2\Big[\frac{\hat{Q}\cdot \hat{Q}}{N}+k_3\frac{(\hat{Q}\times \hat{Q}\times \hat{Q})^{(0)}}{N(N-1)}\Big]\, ,
\end{eqnarray}
where $\hat{n}_d=d^\dag\cdot\tilde{d}$ is the $d$-boson number operator and the $\hat{Q}$ operator is taken as that defined in (\ref{QLII}).
If defining $C_1=\xi$ and $C_2=-(1-\xi)$, the Hamiltonian (\ref{HSC}) will precisely reduce to that employed in \cite{Fortunato2011}.
The classical potential function can be derived as
\begin{eqnarray}\label{VSC}
V_{\,\mathrm{S_C}}(\beta,\gamma)&=&\frac{1}{N}\langle\beta, \gamma,
N|\hat{H}_{\mathrm{S_C}}|\beta, \gamma,N\rangle|_{N\rightarrow\infty}\\ \nonumber
&=&C_1\frac{4\beta^2}{(1+\beta^2)^6}+C_2\frac{4\beta^2}{(1+\beta^2)^2}\\ \nonumber
&\times&\Big[\frac{1}{14}\Big(\chi^2\beta^2-2\sqrt{14}\chi\beta \mathrm{cos}(3\gamma)+14\Big)\\ \nonumber
&+&k_3\frac{\beta}{49\sqrt{5}(1+\beta^2)}\Big(\chi^3\beta^3(2\mathrm{cos}^2(3\gamma)-1)\\ \nonumber
&+&42\chi\beta-\sqrt{14}(14+3\chi^2\beta^2)\mathrm{cos}(3\gamma)\Big)\Big]\, .
\end{eqnarray}
Similarly, a $\mathrm{cos}^2(3\gamma)$ term appears in the potential function, indicating the feasibility of triaxial deformation.
In fact, it has been revealed~\cite{Fortunato2011} that a triaxial minimum for the extended consistent-$Q$ Hamiltonian can indeed manifest at the mean-field level within a tiny parameter region with $\chi\sim-\sqrt{7}/2$.

\begin{center}
\vskip.2cm\textbf{(D). Dynamic triaxiality}
\end{center}\vskip.2cm

\begin{figure}
\begin{center}
\includegraphics[scale=0.25]{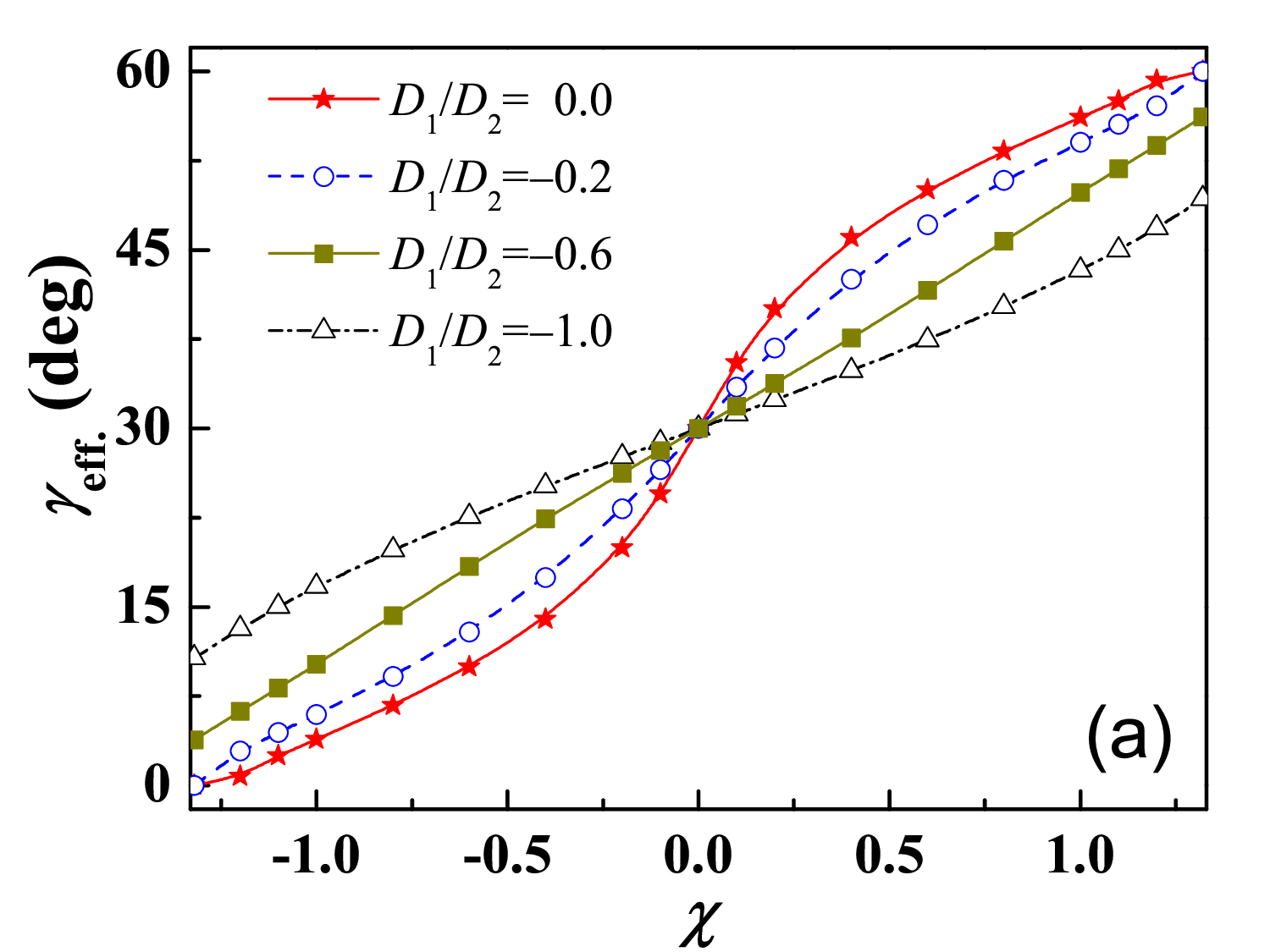}
\includegraphics[scale=0.25]{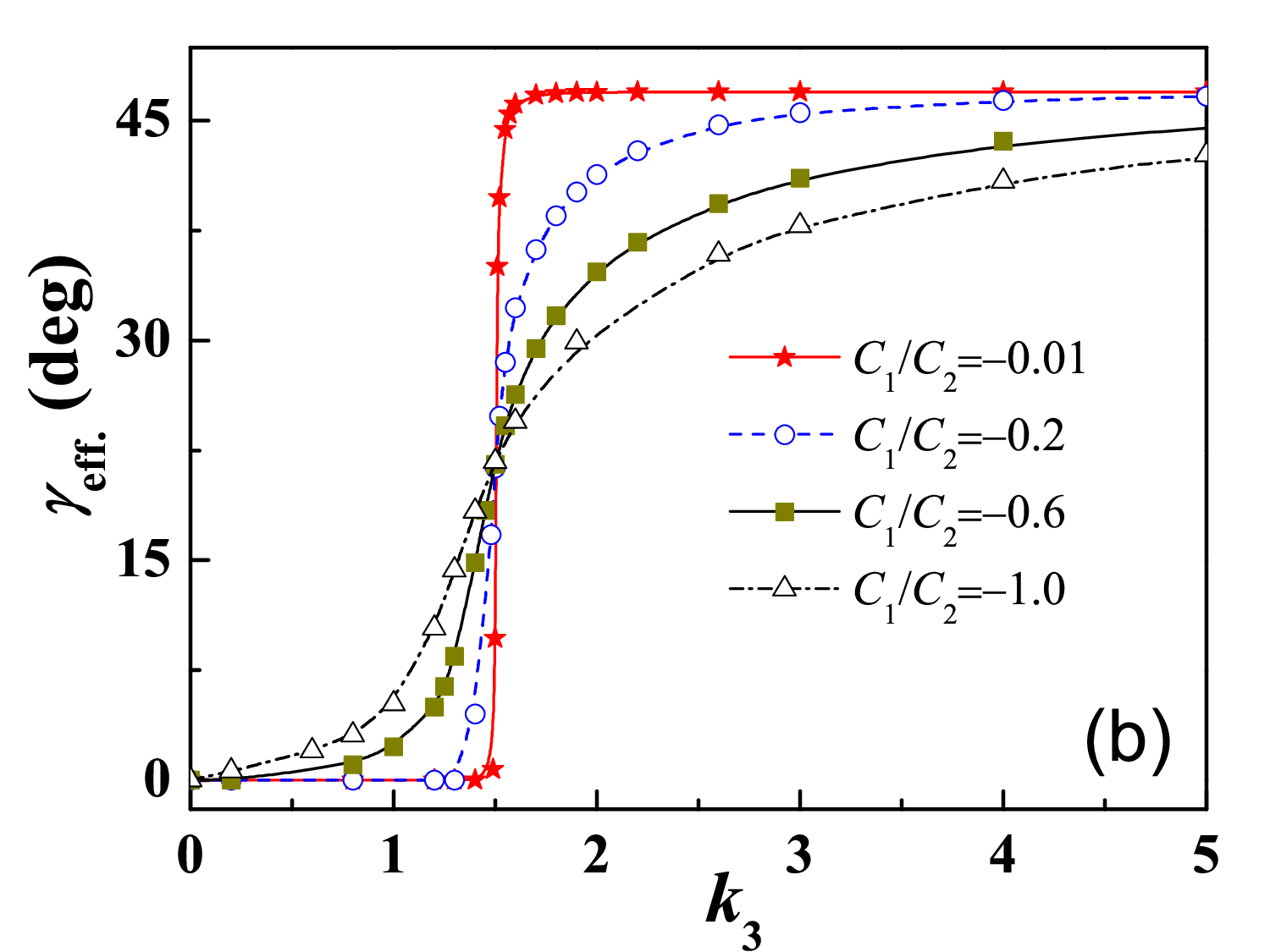}
\caption{(a) The $\gamma_{\mathrm{eff}.}$ values derived from the consistent-$Q$ Hamiltonian (\ref{HSD}) for different ratios of $D_1/D_2$ are shown as a function of $\chi$. (b) Similarly, the results derived from the extended consistent-$Q$ Hamiltonian (\ref{HSC}) for different $C_1/C_2$ are shown as a function of $k_3$. In these calculations, the total number has been fixed at $N=9$.}\label{F02}
\end{center}
\end{figure}

Unlike intrinsic triaxial deformations,
dynamic triaxiality caused by effective triaxial deformations can be identified using the formula (\ref{effgamma}), even when considering an IBM Hamiltonian up to two-body terms.
Since dynamic triaxial deformation, as discussed here, has the same meaning as effective triaxial deformation, we will use these terms interchangeably.
In principle, Eq.~(\ref{effgamma}) can be applied to calculate $\gamma_{\mathrm{eff.}}$ in both $\gamma$-rigid and $\gamma$-soft systems, but the latter cases will be particularly
addressed in the following.
To illustrate the dynamical triaxiality in such situation, we present two examples for $\hat{H}_\mathrm{S}$: the consistent-$Q$ Hamiltonian and its extension defined in (\ref{HSC}).
The consistent-$Q$ Hamiltonian~\cite{Warner1983} can be written as
\begin{eqnarray}\label{HSD}
\hat{H}_\mathrm{S_D}=D_1\hat{n}_d+D_2\frac{\hat{Q}\cdot \hat{Q}}{N}\, ,
\end{eqnarray}
which is simply derived from (\ref{HSC}) by omitting the three-body term.
This Hamiltonian form with $\hat{Q}$ defined in (\ref{QLII}) is frequently employed in the IBM to describe conventional collective modes, which correspond to various dynamical symmetry limits.
Specifically, $\hat{H}_\mathrm{S_D}$ describes the U(5) limit if $D_1>0$ and $D_2=0$; the O(6) limit when $D_1=0$, $D_2<0$ and $\chi=0$; and the SU(3) (or $\overline{\mathrm{SU(3)}}$) limit
if $D_1=0$, $D_2<0$ and $\chi=-\sqrt{7}/2$ (or $\sqrt{7}/2$).
Similarly, one can derive the corresponding classical potential using the coherent state method, which is obtained as
\begin{eqnarray}\label{VSD}
V_{\,\mathrm{S_D}}(\beta,\gamma)&=&\frac{1}{N}\langle\beta, \gamma,
N|\hat{H}_{\mathrm{S_D}}|\beta, \gamma,N\rangle|_{N\rightarrow\infty}\\ \nonumber
&=&D_1\frac{4\beta^2}{(1+\beta^2)^6}+D_2\frac{2\beta^2}{7(1+\beta^2)^2}\\ \nonumber
&\times&\Big[\chi^2\beta^2-2\sqrt{14}\chi\beta \mathrm{cos}(3\gamma)+14\Big]\, .
\end{eqnarray}
Clearly, the term dependent on $\gamma$ is solely proportional to $\mathrm{cos}(3\gamma)$, which means that the consistent-$Q$ Hamiltonian cannot induce a stable triaxial deformation at the mean-field level. Instead, a nonzero $\gamma_{\mathrm{eff}.}$, which signifies effective triaxial deformation, can be obtained using Eq.~(\ref{effgamma}). To see how the effective deformation varies with model parameters, we solve the values of $\gamma_{\mathrm{eff}.}$ from the consistent-$Q$ Hamiltonian ($N=9$) within the context of the SU(3)-O(6)-$\overline{\mathrm{SU(3)}}$ transition. The results as a function of $\chi$ are presented in Fig.~\ref{F02}(a). This type of model evolution has traditionally been employed to describe the prolate to oblate shape transition~\cite{Jolie2001}. When utilizing the extended consistent-$Q$ Hamiltonian defined in (\ref{HSC}),
this shape transition can alternatively be characterized by model evolution as a function of $k_3$. To provide a parallel comparison, we also compute values of $\gamma_{\mathrm{eff}.}$ based on the Hamiltonian given in (\ref{HSC}) with $\chi=-\sqrt{7}/2$ and present the results as a function of $k_3$ in Fig.~\ref{F02}(b).

As illustrated in Fig.~\ref{F02}(a), $\gamma_{\mathrm{eff}.}$ along the precise SU(3)-O(6)-$\overline{\mathrm{SU(3)}}$ transitional routine ($D_1/D_2=0$) exhibits a monotonic increase from $0^\circ$ to $60^\circ$ with the O(6) point ($\chi=0$) maintaining at $\gamma_{\mathrm{eff}.}=30^\circ$. This description aligns with the usual interpretation of the SU(3), O(6) and $\overline{\mathrm{SU(3)}}$ limits, which corresponds to prolate, $\gamma$-unstable and oblate shapes, respectively. Note that the potential function given in (\ref{VSD}) suggests that the O(6) point corresponds to a $\gamma$-unstable configuration at the mean-field level but with the average value of $\langle{\gamma_0}\rangle=30^\circ$, which is consistent with the effective deformation, $\gamma_{\mathrm{eff}.}=30^\circ$, obtained here. Regarding this point, we hope to mentioned that for finite-$N$ boson number systems, the $\gamma$-unstable state can be generated from the rigid triaxial configuration with $\gamma=30^\circ$, as analyzed in \cite{Otsuka1987}, which establishes the equivalence between two descriptions of triaxiality in nuclei. Furthermore, a rapid change in $\gamma_{\mathrm{eff}.}$ is shown to occur near the O(6) point, suggesting that the evolution depicted in Fig.~\ref{F02}(a) represents a finite-$N$ precursor~\cite{Jolie2001} of the prolate-oblate shape phase transition defined in the large-$N$ limit~\cite{Iachellobook}. Additionally, it is shown that incorporating the U(5) component will smooth out the transitional feature as indicated by the results for $D_1/D_2\neq0$.

The results in Fig.~\ref{F02}(b) regarding the extended consistent-$Q$ Hamiltonian indicate a significantly sharper prolate-oblate shape phase transitional behavior. This observation persists despite
the smoothing effects introduced by incorporating the U(5) components ($C_1/C_2\neq0$). In fact, it has been demonstrated~\cite{Zhang2012} that this transition in the SU(3) limit ($C_1/C_2=0$) is a first-order quantum phase transition due to the levels crossing even at finite boson numbers. In the SU(3) limit, effective $\gamma$ deformations as a function of $k_3$ are restricted to only two values, which become degenerated at the critical point of the prolate-oblate phase transition. For instance, the case with $N=9$ may give $\gamma_{\mathrm{eff}.}=0^\circ$ in the prolate phase corresponding to the SU(3) IRREP $(\lambda,\mu)=(18,0)$ and $\gamma_{\mathrm{eff}.}\approx47^\circ$ in the oblate phase associated with $(\lambda,\mu)=(2,8)$. To circumvent such extreme situation, we have opted not to consider the $C_1/C_2=0$ case in Fig.~\ref{F02}(b) and instead used $C_1/C_2=-0.01$ to signify the similar situation.
It should be mentioned that the $\gamma_{\mathrm{eff}.}$ values obtained from the SU(3) limit consist well with the relation
\begin{equation}\label{su3gamma}
\gamma=\mathrm{tan}^{-1}\Big(\frac{\sqrt{3}(\mu+1)}{2\lambda+\mu+3}\Big)\,
\end{equation}
proposed in \cite{Leschber1987}. A notable example is that the effective $\gamma$ deformation in the SU(3) oblate phase is approximately $\gamma_{\mathrm{eff}.}\approx60^\circ$ for $N=8$, as opposed to $\gamma_{\mathrm{eff}.}\approx47^\circ$ for $N=9$ as exhibited in Fig.~\ref{F02}(b). Similar results can also be derived using Eq.~(\ref{su3gamma}), as the maximal oblate IRREP in the SU(3) limit will change from $(\lambda,\mu)=(2,8)$ for $N=9$ to $(\lambda,\mu)=(0,8)$ for $N=8$.

\begin{figure}
\begin{center}
\includegraphics[scale=0.175]{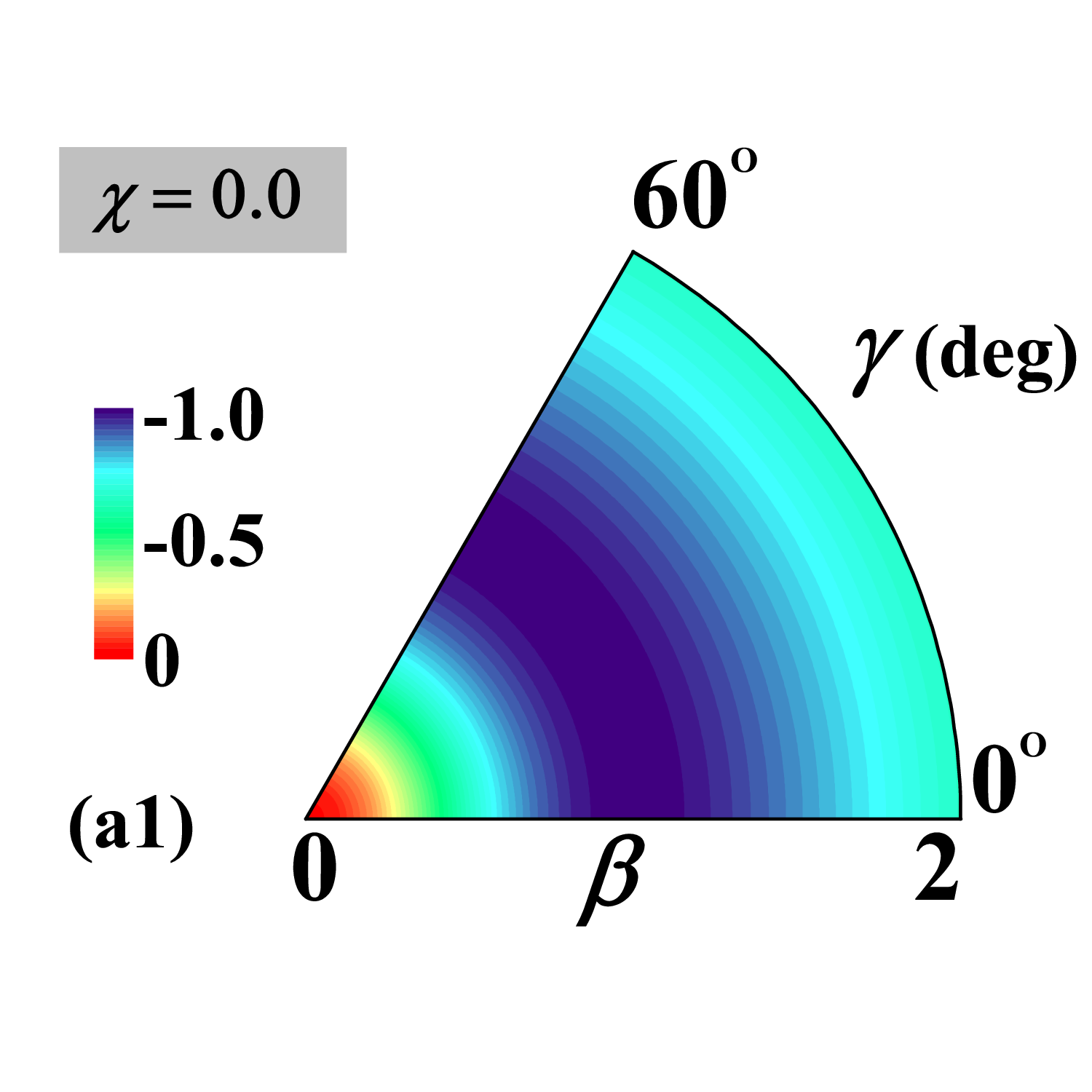}
\includegraphics[scale=0.175]{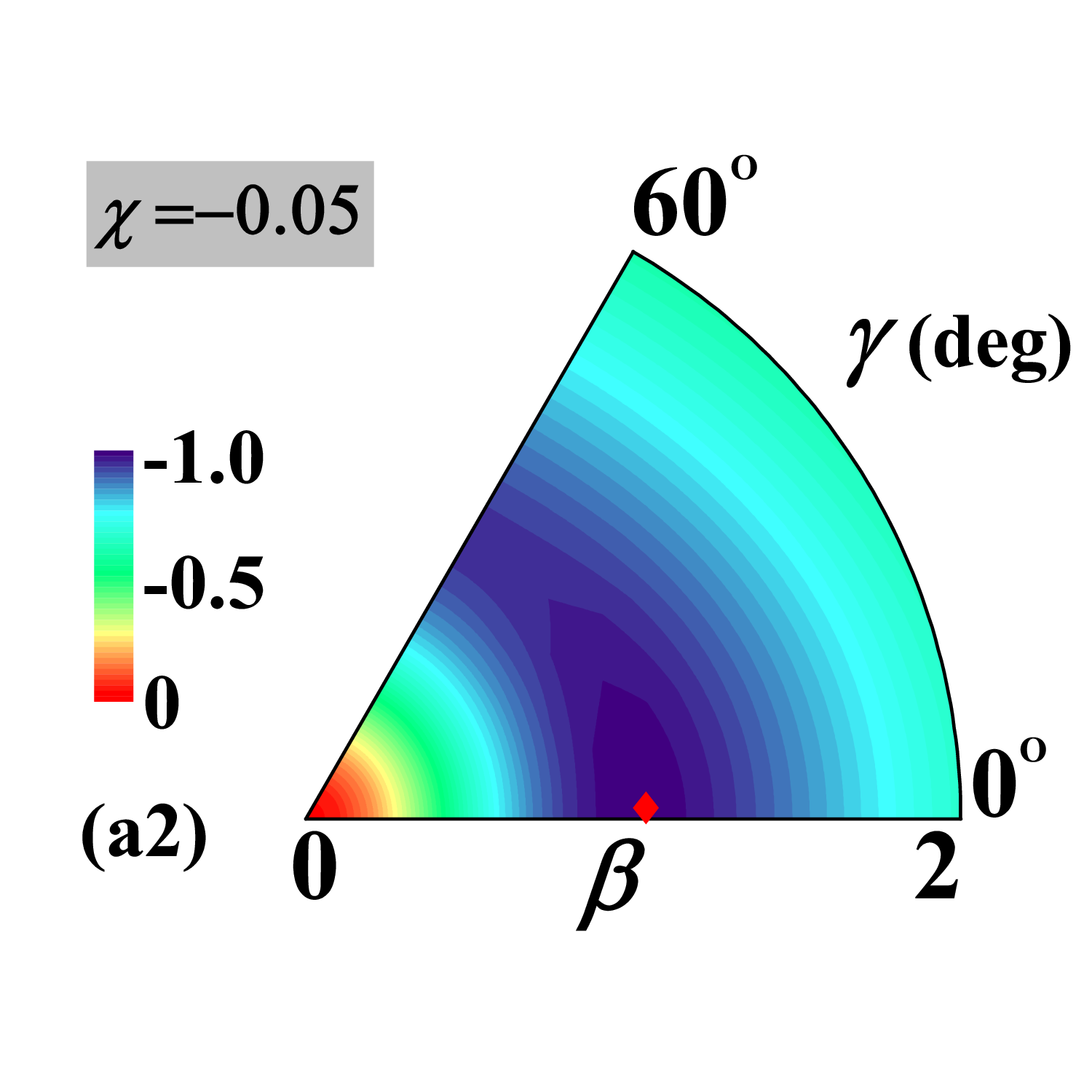}
\includegraphics[scale=0.175]{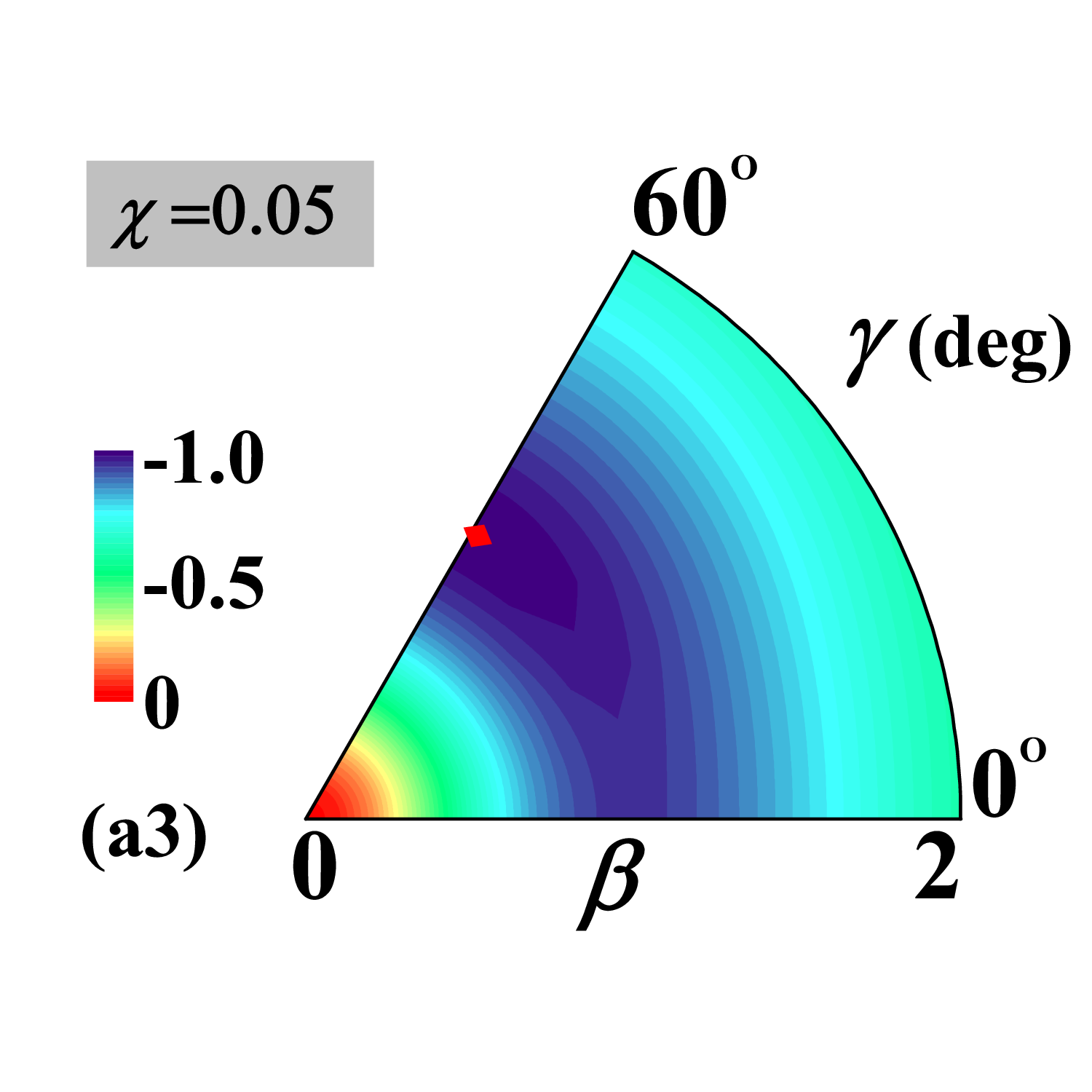}
\caption{(a1) The O(6) potential surface derived from Eq.~(\ref{VSD}) with $\chi=0$, $D_1=0$ and $D_2=-1$.
(a2) The same as in (a1) but for $\chi=-0.05$ with the minimal point $(\beta_0,~\gamma_0)\simeq(1.0,~0^\circ)$ characterized by a diamond (red) symbol.(a3) The same as in (a2) but for $\chi=0.05$ with the minimal point locating at $(\beta_0,~\gamma_0)\simeq(1.0,~60^\circ)$.}\label{F03}
\end{center}
\end{figure}
In cases deviating from the SU(3) limit, $\gamma_{\mathrm{eff}.}$ is usually close to the $\gamma_0$ value obtained through the coherent state method, provided that the $\gamma$ deformation can be well identified at the mean-field level. For instance, if taking $(B_1,~B_2)=(-30,~180)$ for $\hat{H}_{\mathrm{S_B}}$ as proposed in the scheme (B), one can derive the mean-filed deformation $\gamma_0\approx30^\circ$, and the corresponding effective $\gamma$ deformation in this case is given by $\gamma_{\mathrm{eff}.}\approx30^\circ$. In contrast, if the mean-field $\gamma$ deformation exhibits significant softness, a substantial difference between these two approaches may arise. For example, one can observe in Fig.~\ref{F03} that the potential at the O(6) point ($\chi=0$) deduced from Eq.~(\ref{VSD}) is indeed $\gamma$ unstable, while the cases with $\chi$ slightly deviating away from the O(6) limit will yield either $\gamma_0=0^\circ$ or $\gamma_0=60^\circ$ along with a minimal valley, indicating $\gamma$ softness at the mean-field level.
Nevertheless, the results shown in Fig.~\ref{F02}(a) manifest that effective $\gamma$ deformations in the corresponding cases ($D_1/D_2=0$) with different $\gamma_0$ may all give $\gamma_{\mathrm{eff}.}\approx30^\circ$. It means that calculation of $\gamma_{\mathrm{eff}.}$ utilizing ground-state wave function has taken into account the fluctuation in $\gamma$.

\begin{center}
\vskip.2cm\textbf{IV. Results and discussions}
\end{center}\vskip.2cm

\begin{figure*}
\begin{center}
\includegraphics[scale=0.21]{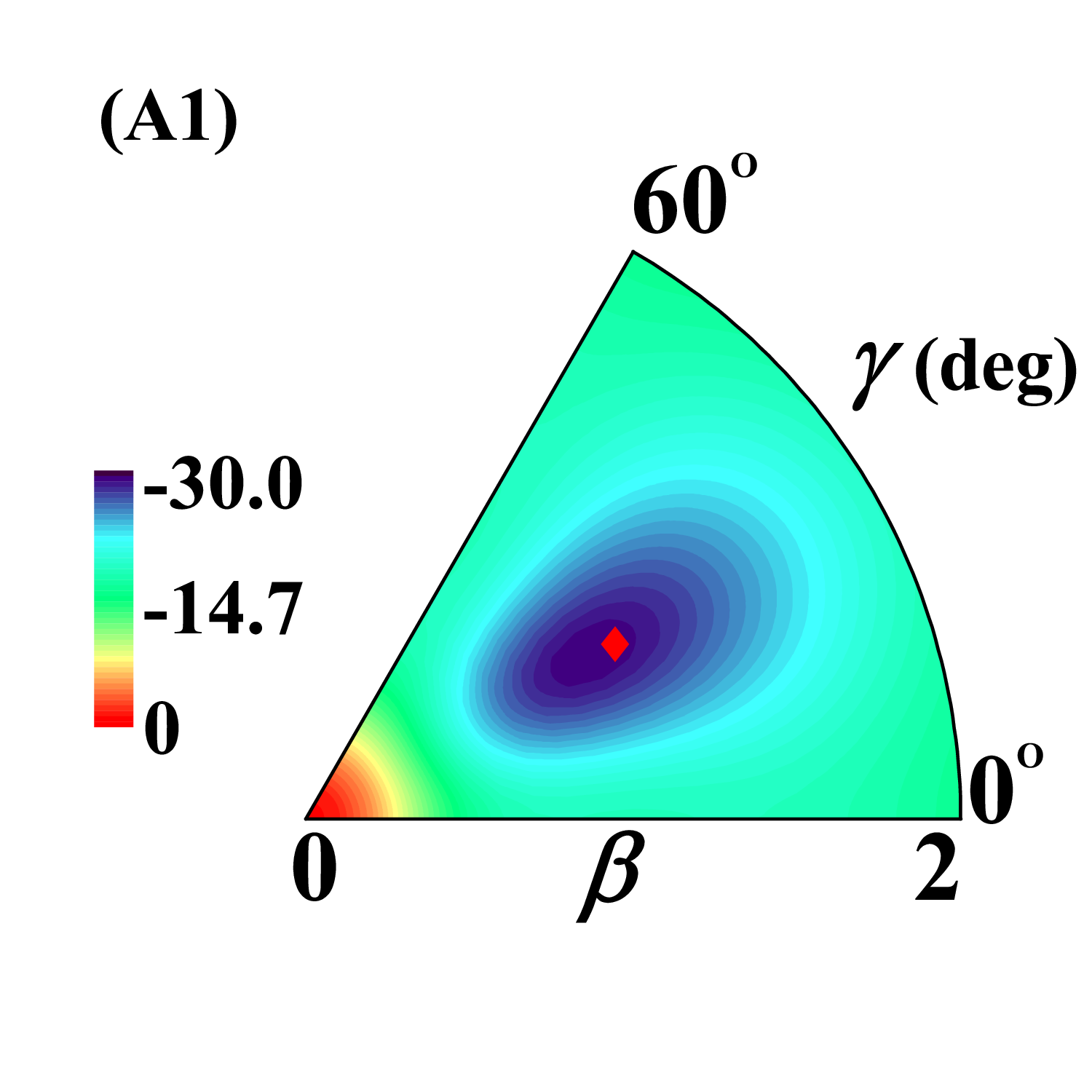}
\includegraphics[scale=0.21]{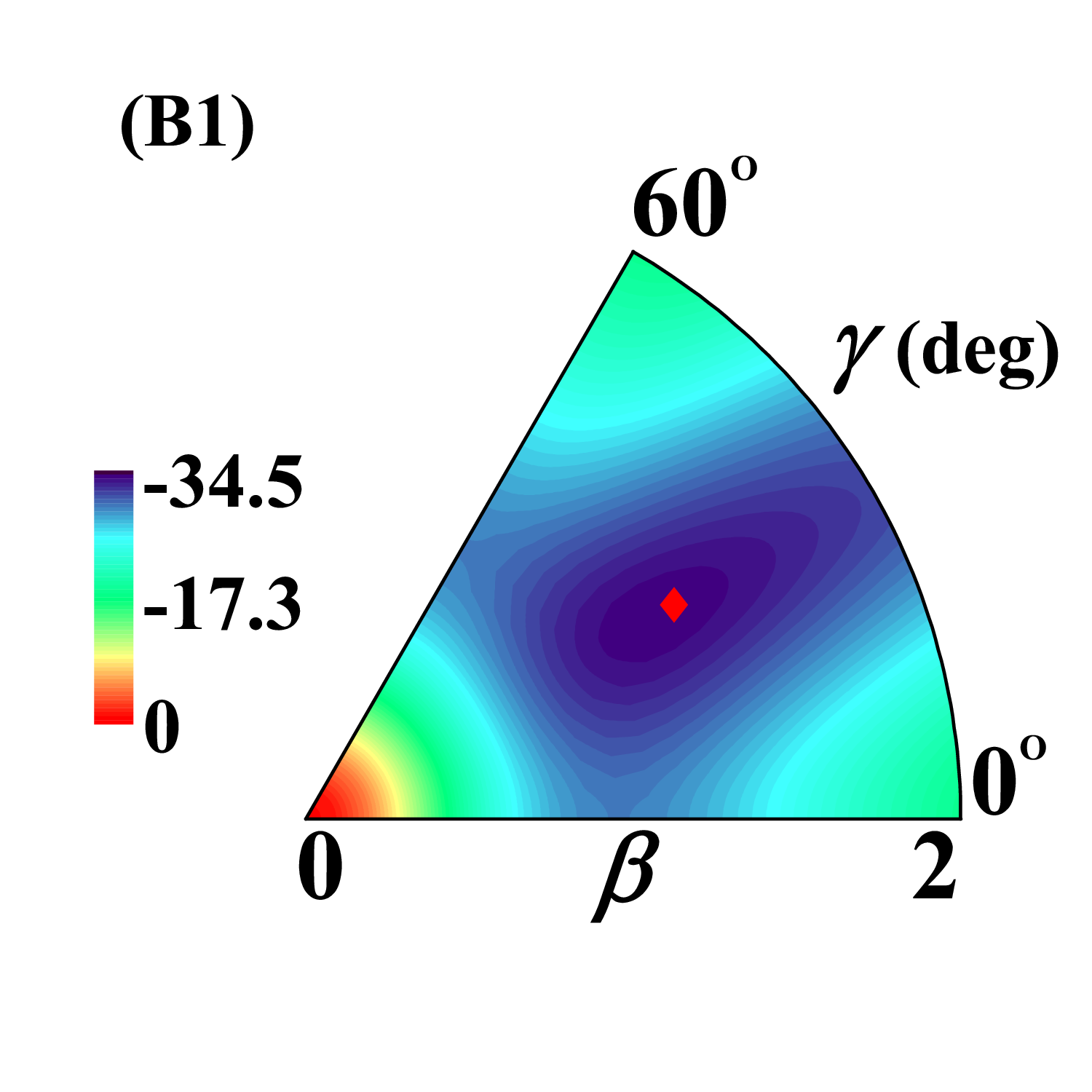}
\includegraphics[scale=0.21]{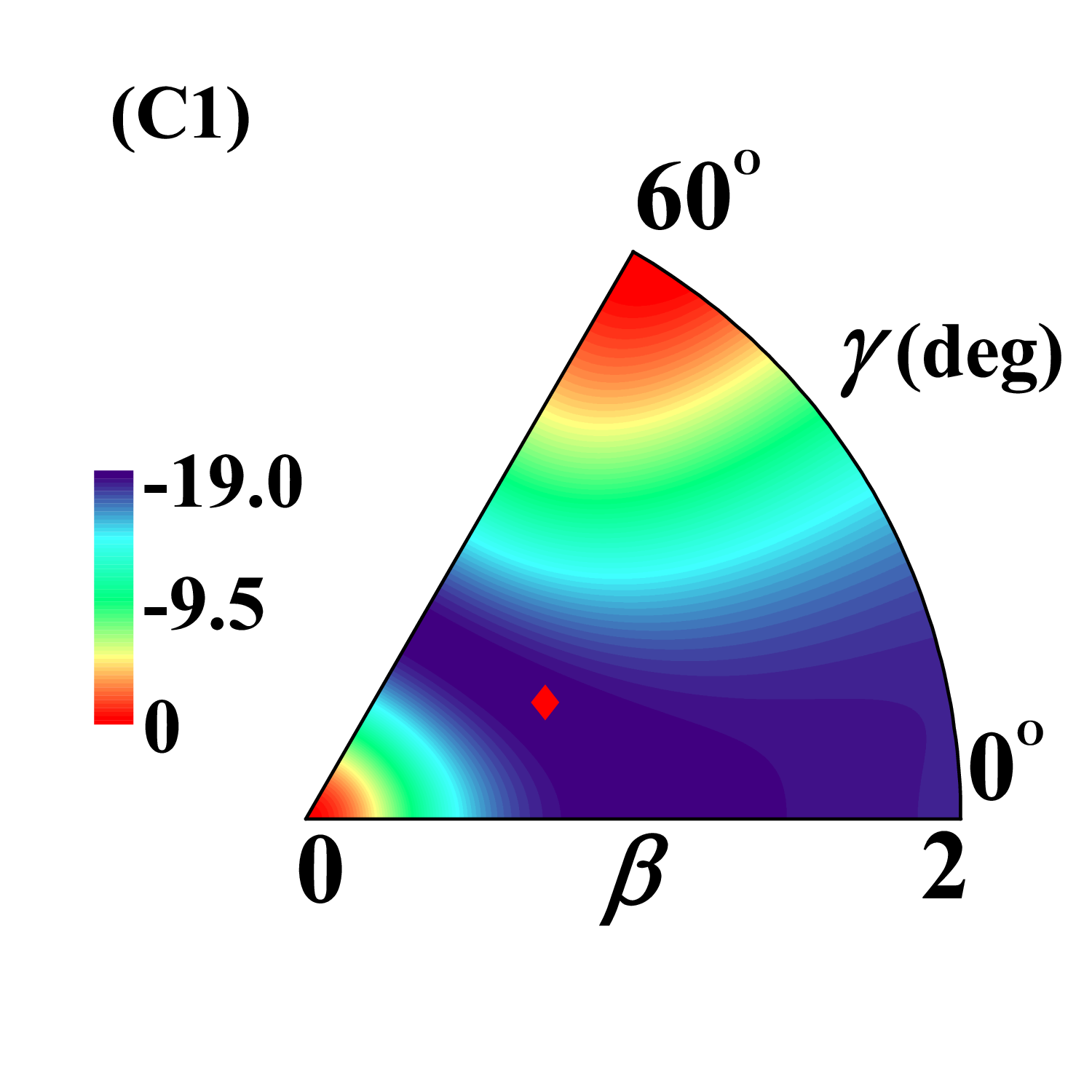}
\includegraphics[scale=0.2]{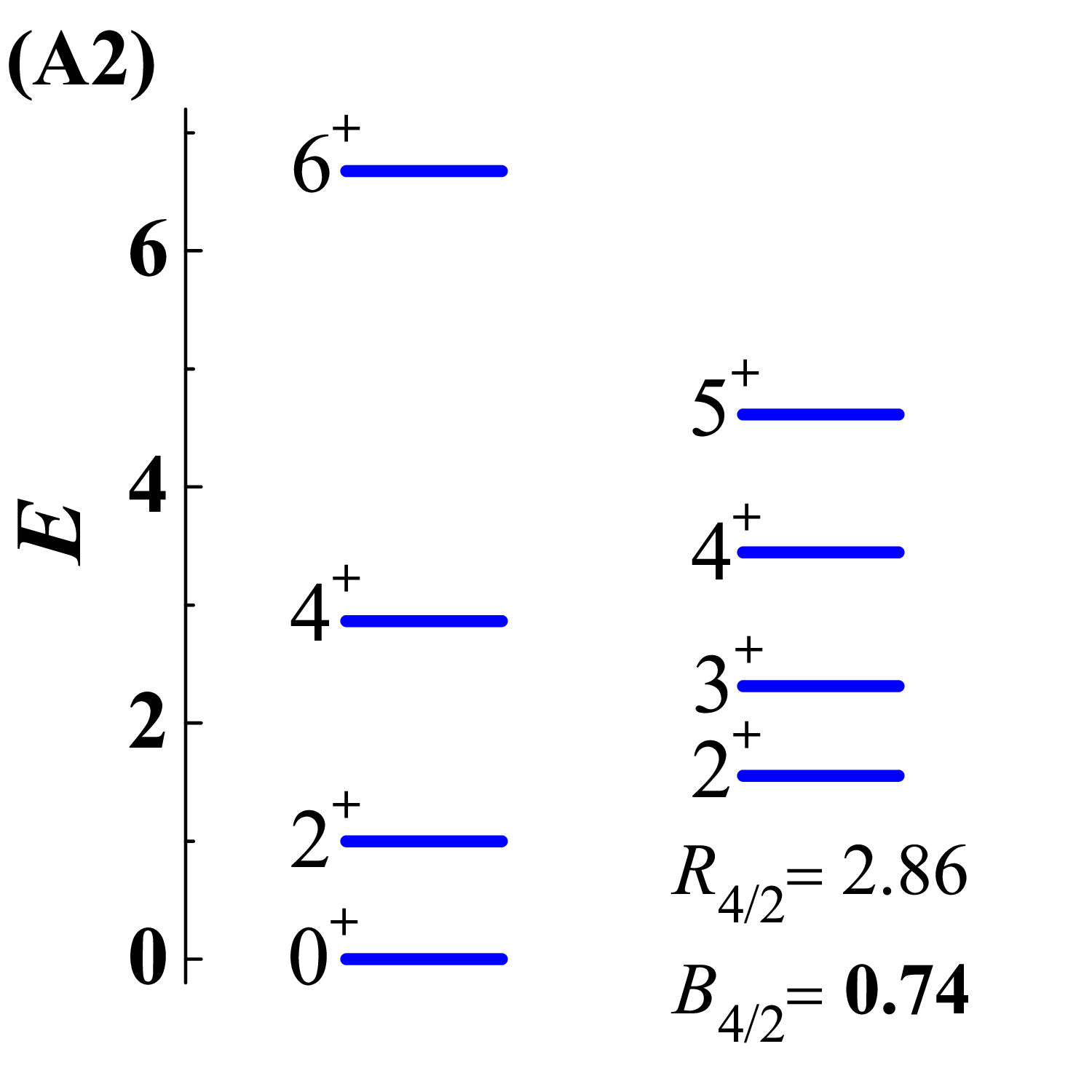}
\includegraphics[scale=0.2]{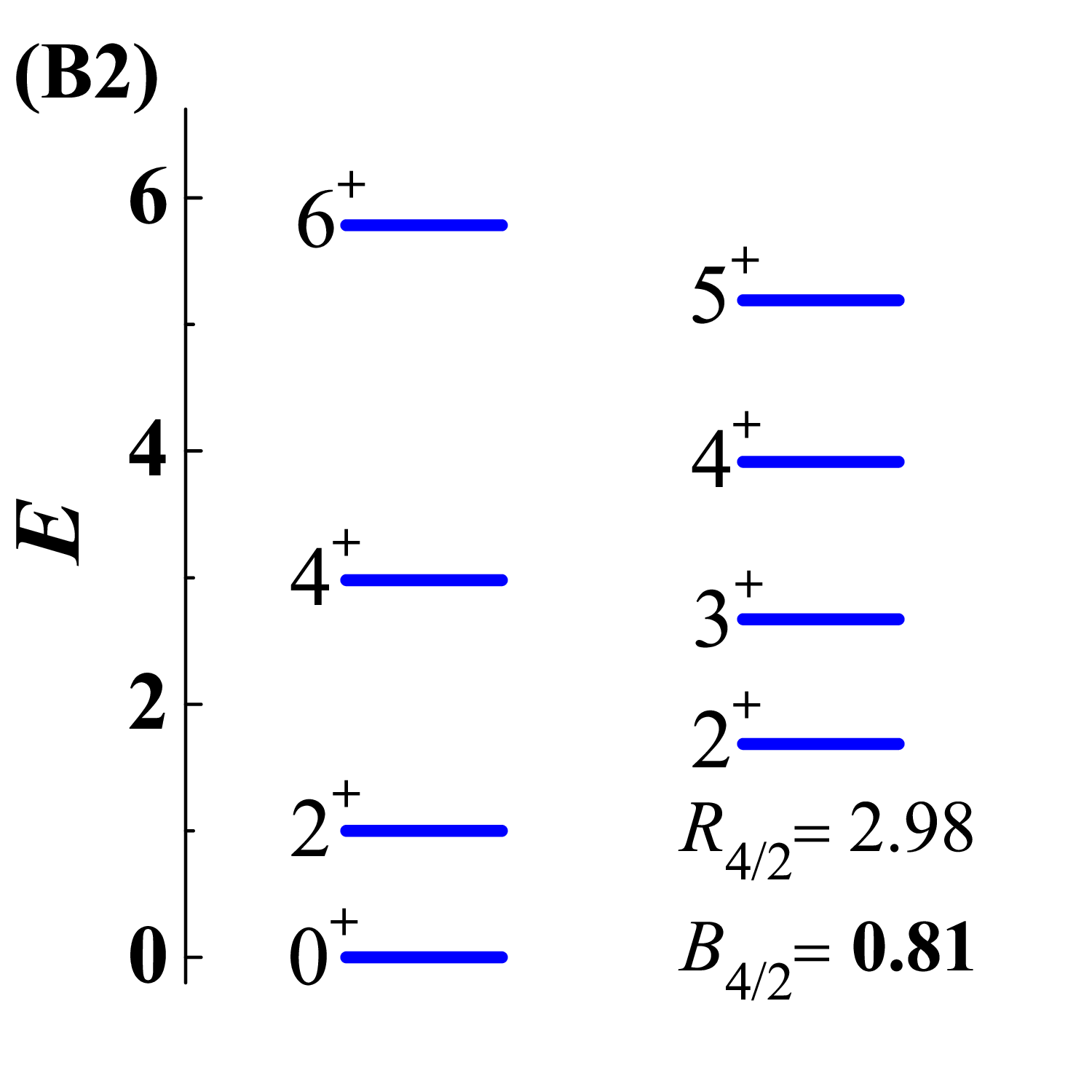}
\includegraphics[scale=0.2]{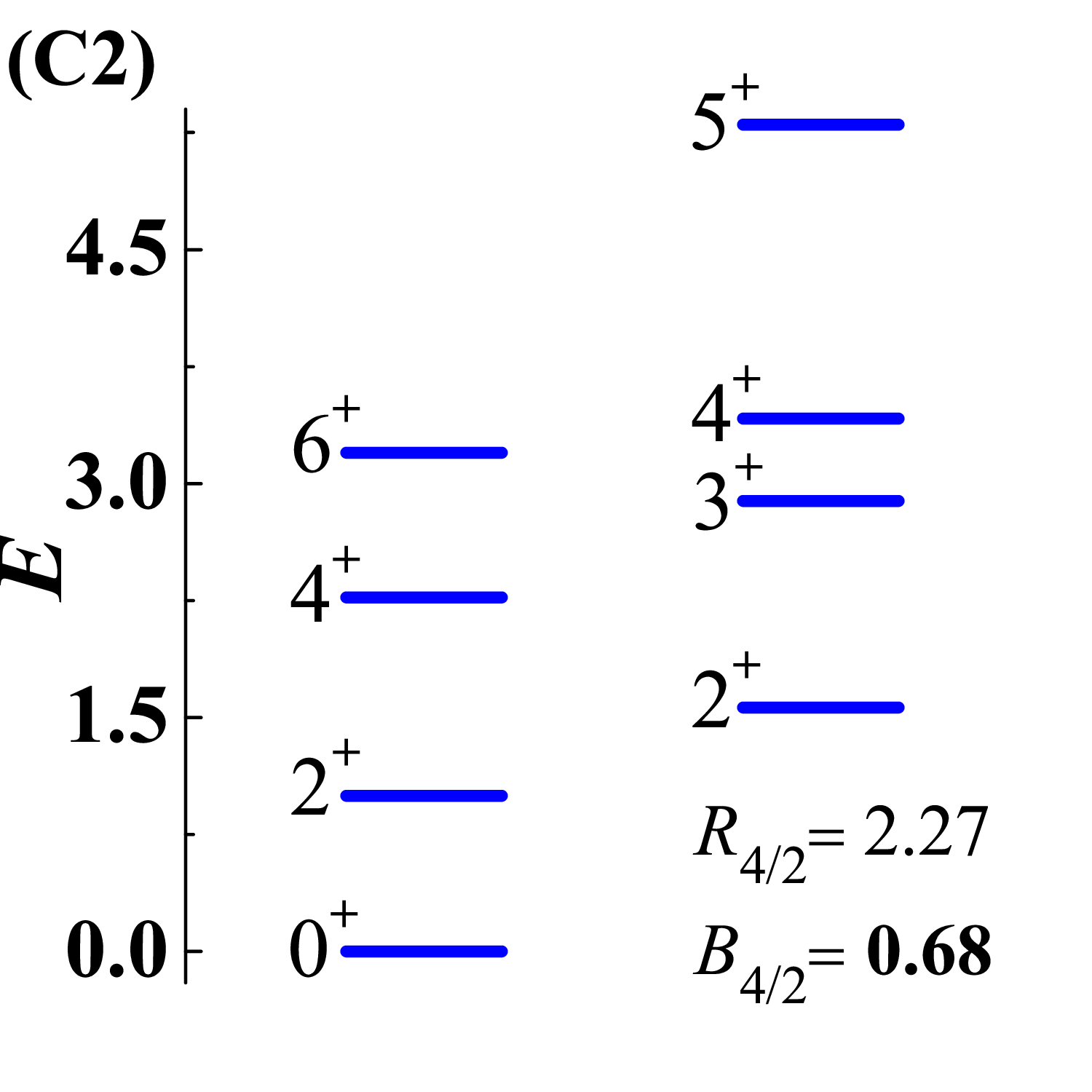}
\caption{(A1) The potential surface derived from the scheme (A) with the minimal point $(\beta_0,~\gamma_0)\approx(1.0,~30^\circ)$ characterized by a diamond (red) symbol.
(B1) The potential surface derived from the scheme (B) with the minimal point $(\beta_0,~\gamma_0)\approx(1.3,~30^\circ)$. (C1) The potential surface derived from the scheme (C) with the minimal point $(\beta_0,~\gamma_0)\approx(0.8,~30^\circ)$.
(A2) The low-lying level pattern for $N=9$, derived from the triaxial Hamiltonian based on the scheme (A), is presented with all level energies
being normalized to $E(2_1^+)=1.0$. (B2) The same as in (A2) but for the result obtained based on the scheme (B). (C2) The same as in (A2) but for the result obtained based on the scheme (C). All the parameters adopted in the calculations are illustrated in the text. }\label{F1}
\end{center}
\end{figure*}

As discussed above, one can choose different $\hat{H}_\mathrm{S}$ to generate a triaxial deformation, from which $\hat{H}_\mathrm{D}$ associate with a specific rotor mode can be derived using two approaches to construct the full triaxial Hamiltonian. In the following, we will exemplify that the $B(E2)$ anomaly features produced from an asymmetric top, as observed in Fig.~\ref{F0}, can similarly be generated by the triaxial rotor modes developed through different schemes. To calculate $B(E2)$ transitions, the transitional operator is adopted as
\begin{eqnarray}
\hat{T}^{E2}=e_\mathrm{B}\hat{Q}_u\, ,
\end{eqnarray}
in which $e_\mathrm{B}$ represents the effective charge and the quadrupole operator $\hat{Q}_u$ is taken as same as that involved in the Hamiltonian.
It should be emphasized that the present analysis aims primarily to illustrate the correlation between $B(E2)$ anomaly and triaxility in the IBM rather than to tell which triaxial scheme is more reasonable in theory. Here, triaxiality is defined as arising from both mean-field triaxial deformations and effective triaxial deformations, which means both the $\gamma$-rigid and $\gamma$-soft cases all under this definition.

\begin{center}
\vskip.2cm\textbf{(a). $B(E2)$ anomaly in different triaxial schemes}
\end{center}\vskip.2cm

First, we analyze the results derived from the triaxial Hamiltonian utilizing the mean-field approach. Specifically, the parameters (in any units) for $\hat{H}_\mathrm{S_A}$, $\hat{H}_\mathrm{S_B}$ and $\hat{H}_\mathrm{S_C}$ as defined in Section III are chosen as $(A_1,~A_2,~A_3,~A_4)=(-30,~0.0,~-3.0,~230)$, $(B_1,~B_2)=(-30,~180)$ and $(C_1,~C_2,~k_3,~\chi)=(2.5,~-30,~1.95,~-1.323)$, respectively. It is shown that the coefficients in front of the $\hat{Q}\cdot\hat{Q}$ term have been assigned identical value across all three schemes in order to generate classical potentials of comparable scale. With these parameters, the mean-field deformations are obtained as $(\beta_0,~\gamma_0)\approx(1.0,~30^\circ),~(1.3,~30^\circ)$ and $(0.8,~30^\circ)$, respectively, indicating that all the three cases exhibit intrinsic triaxial deformations.
Employing the aforementioned mean-field procedure allows further determination of the parameters involved in $\hat{H}_\mathrm{D}$, resulting in $(a,~b,~c)=(3.0,~0.56,~-0.023),~(3.0,~0.58,~-0.025)$ and $(3.5,~0.67,~-0.06)$ for the schemes (A), (B) and (C). To evaluate $(a,~b,~c)$ through the mapping formulas (\ref{abc}), the inertia parameters $\Gamma_\alpha$ are assumed to follow $\Gamma_1:\Gamma_2:\Gamma_3=3:1:4$, consistent with those used for the asymmetric top depicted in Fig.~\ref{F0}. Once all the parameters are established, one can diagonalize the full Hamiltonian $\hat{H}_{\mathrm{Tri}}$ to derive level energies and $B(E2)$ values. The calculated level patterns ($N=9$) along with their corresponding potential surfaces are presented in Fig.~\ref{F1}.

As seen in Fig.~\ref{F1}(A1)-(C1), triaxial deformation can be readily identified from the potential surface in each case. By contrast, the triaxial minima observed in the cases (A) and (B) exhibit greater stability compared to that generated in the case (C), where a narrow equilibrium valley extends from prolate to oblate. This feature has also been noted in the previous mean-field analysis~\cite{Fortunato2011} of the extended consistent-$Q$ Hamiltonian, indicating that a soft triaxial deformation occurs accompanied by the prolate-oblate shape phase transition. As depicted in Fig.~\ref{F1}(A2)-(C2), the different cases reveal varying odd-even staggering in their $\gamma$-band levels. According to the analysis in \cite{Zamfir1991,McCutchan2007}, a $\gamma$-unstable structure or extremely $\gamma$ soft triaxial structure is expected to exhibit $\gamma$-band levels clustering as $(2_\gamma^+), (3_\gamma^+,~4_\gamma^+),\cdots$, whereas a $\gamma$-rigid triaxial structure clustering in the form of $(2_\gamma^+,~3_\gamma^+), (4_\gamma^+,~5_\gamma^+),\cdots$. Then, one may determine that a relatively soft triaxial structure is presented in Fig.~\ref{F1}(C2). Such a result is actually in accord with the mean-field pictures shown in Fig.~\ref{F1}, where it can be observed that the potential depicted in panel (C1) remains notably soft in $\gamma$, despite the presence of a shallow minimum ($\beta_0,\gamma_0$) that can be still identified numerically. In addition, all three cases have developed a low-energy $\gamma$ band adjacent to the yrast band. In particular, an unconventional collective feature with $R_{4/2}>2.0$ and $B_{4/2}<1.0$ emerges across all schemes. This suggests that the $E2$ anomalous behavior previously identified within the SU(3) model~\cite{Zhang2022} is theoretically a more robust collective phenomenon.

The results presented in Fig.~\ref{F1}(A1) and Fig.~\ref{F1}(A2) can be also utilize to address the question raised in a recent study~\cite{Wang2023} that the O(6) Hamiltonian, incorporating high-order terms akin to those employed in the SU(3) model~\cite{Zhang2022}, cannot account for $B(E2)$ anomaly in yrast states. As indicated in Eq.~(\ref{VSA}), triaxial deformation cannot be established at the mean-field level using the O(6) Hamiltonian limited to the fourth order terms applied in \cite{Wang2023}. Instead, one can construct a triaxial Hamiltonian based on the O(6) scalar polynomial extended up to the sixth-order terms, as demonstrated here. Consequently, a triaxial spectrum with $B_{4/2}<1.0$ will emerge, as illustrated in Fig.~\ref{F1}(A2). Another point to address is that there is no way to completely separate the intrinsic part from the dynamical part within the schemes under discussions. This remains true even though the full triaxial Hamiltonian in (\ref{Tri}) has been divided into two parts in order to better understand the components related to triaxial rotor. In fact, $\hat{H}_\mathrm{S}$ in each scheme may contribute not only to the band head but also to rotational members of a given band. This indicates that the spectrum generated by $\hat{H}_{\mathrm{Tri}}$ may encompass contributions from modes beyond the triaxial rotor, leading to substantial differences across different triaxial schemes when employing the same mapping procedure.
Nonetheless, the results given in Fig.~\ref{F1} suggest that the low-lying spectra associated with relatively $\gamma$-rigid classical potentials bear similarities to those produced by the rigid rotor model. As observed in Fig.~\ref{F0}, the rotor Hamiltonian (\ref{Hr}) with $\Gamma_1:\Gamma_2:\Gamma_3=3:1:4$ may generated $B_{4/2}\approx1.04$ along with $R_{4/2}\approx2.80$ at $\gamma_\mathrm{e}=30^\circ$. These values are comparatively closer to the ones shown in Fig.~\ref{F1}(A2) and Fig.~\ref{F1}(B2), particularly when contrasted with those displayed in Fig.~\ref{F1}(C2).

Further numerical calculations suggest that the $B_{4/2}<1.0$ features in all the cases as shown in Fig.~\ref{F1} can be eliminated by increasing the boson number $N$ while keeping the other parameters in $\hat{H}_\mathrm{S}$ unchanged. For instance, the ratio in the scheme (A) will rise from $B_{4/2}\approx0.74$ for $N=9$ to $B_{4/2}\approx1.0$ for $N=12$, and further to $B_{4/2}\approx1.05$ for $N=20$. This trend somewhat reflects the finite-$N$ effects on $B(E2)$ anomaly, similar to the cases discussed in the SU(3) model~\cite{Zhang2022}, where these effects are illustrated more clearly based on the analytic relationship between the $\gamma$ deformation and the $N$-dependent triaxial IRREPs $(\lambda,\mu)$ as indicated by Eq.~(\ref{su3gamma}). It should be mentioned that the finite-$N$ corrections to $B(E2)$ transitions may universally manifest in the IBM calculations, and analytical $N$-dependent results can even be derived for cases in the symmetry limits~\cite{Iachellobook}. However, no instances of $B_{4/2}<1.0$ were found to occur within the conventional collective modes, even when accounting for the finite-$N$ effects.

\begin{center}
\vskip.2cm\textbf{(b). $B(E2)$ anomaly with dynamical triaxiality}
\end{center}\vskip.2cm

\begin{figure}
\begin{center}
\includegraphics[scale=0.25]{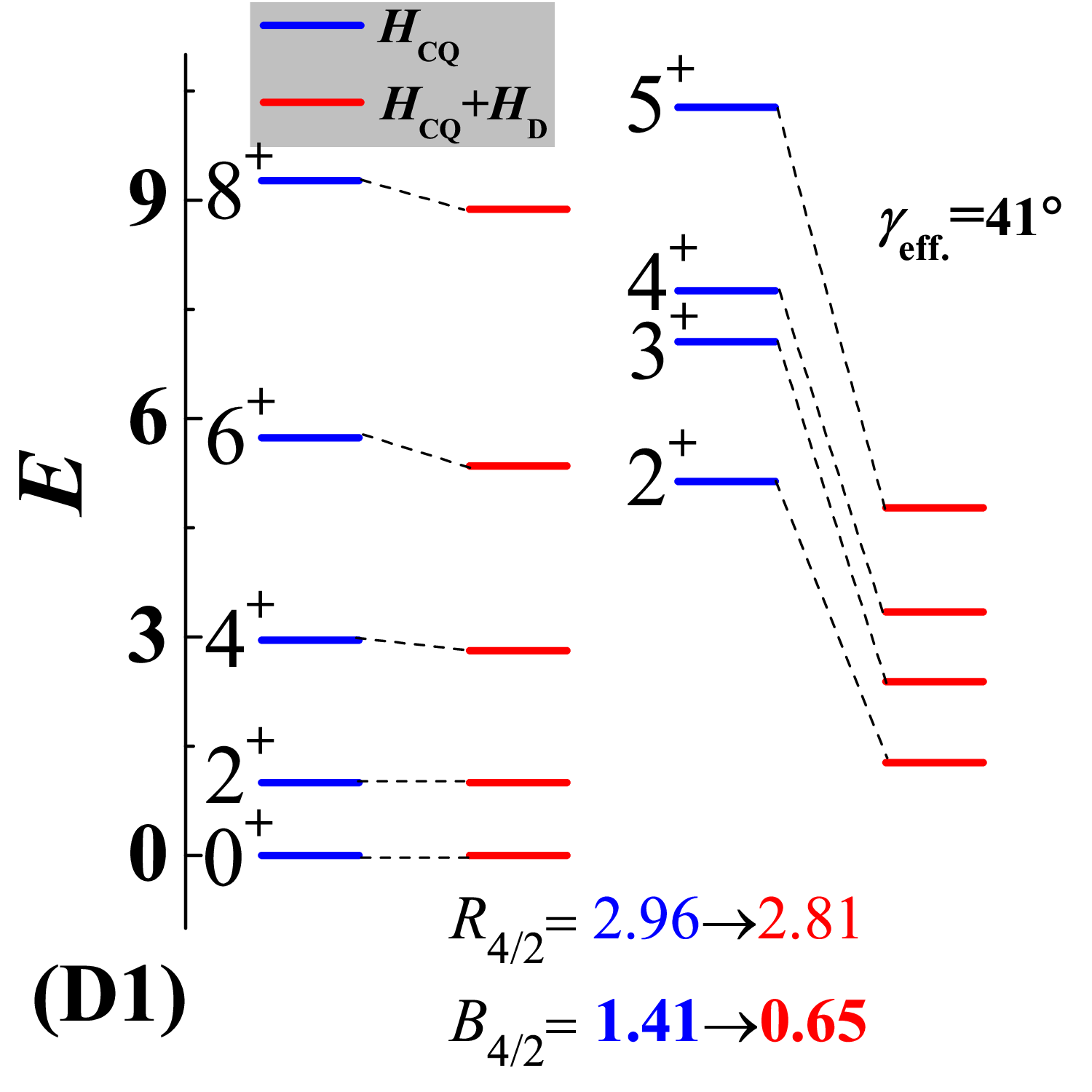}
\includegraphics[scale=0.25]{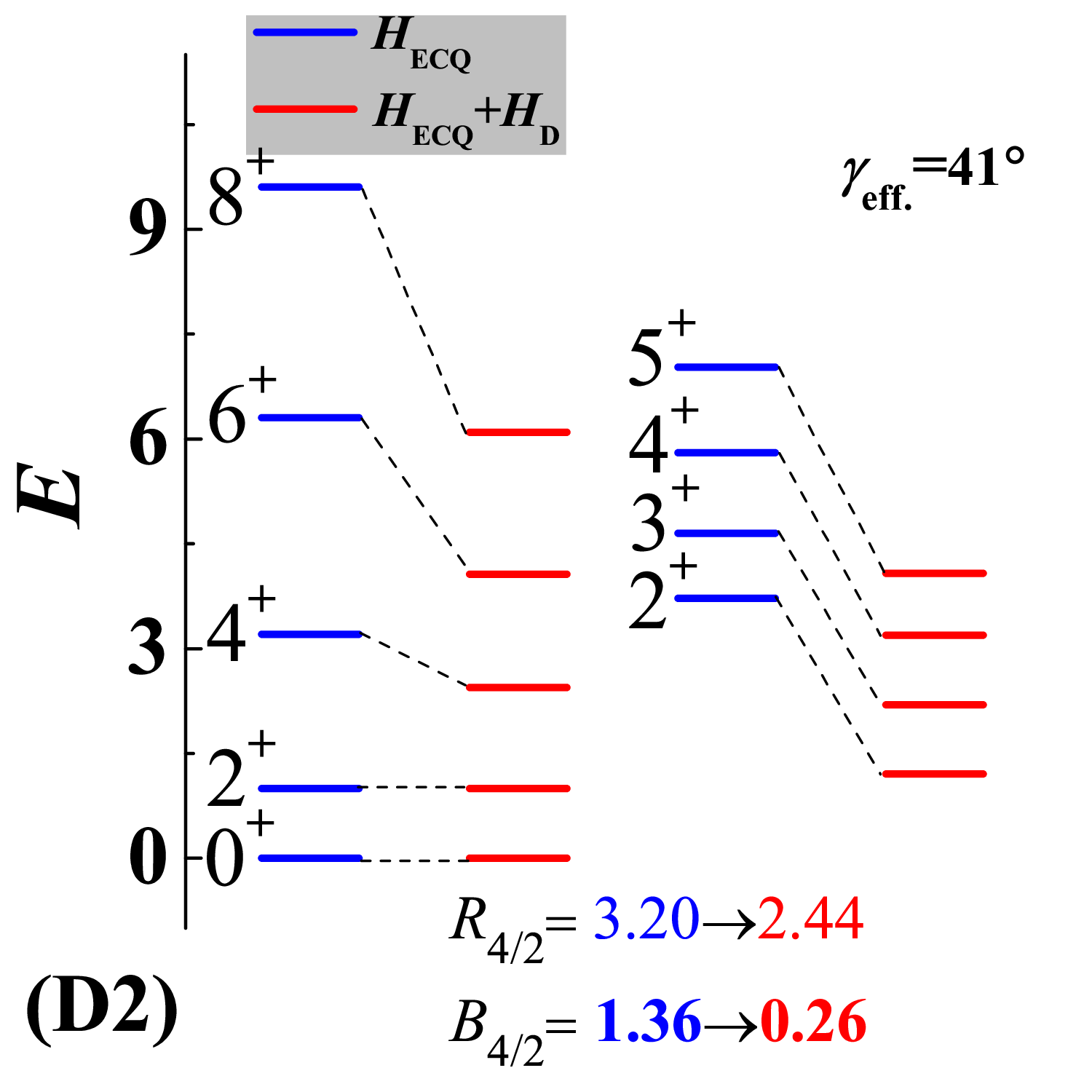}
\caption{(D1) The low-lying level patterns ($N=9$) derived from the consistent-$Q$ Hamiltonian ($H_{\mathrm{CQ}}$) as well as the full triaxial Hamiltonian ($H_{\mathrm{CQ}}+H_{\mathrm{D}}$). (D2) The same as in (D1) but for the results obtained based on the extended consistent-$Q$ formula. The level energies depicted in each panel have been normalized to $E(2_1^+)=1.0$, and all the parameters utilized in calculations are detailed in the text. }\label{F2}
\end{center}
\end{figure}

\begin{figure}
\begin{center}
\includegraphics[scale=0.28]{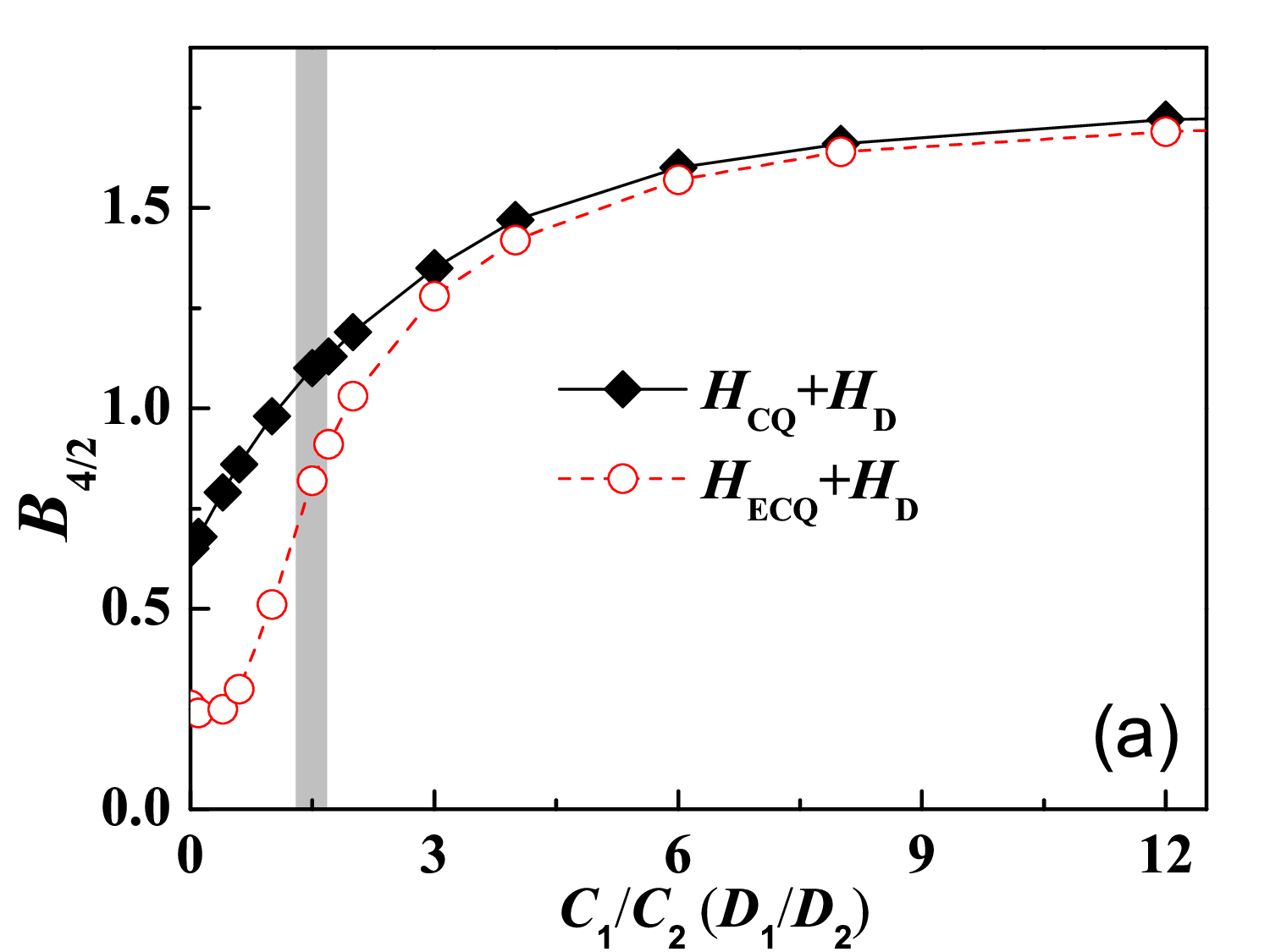}
\includegraphics[scale=0.28]{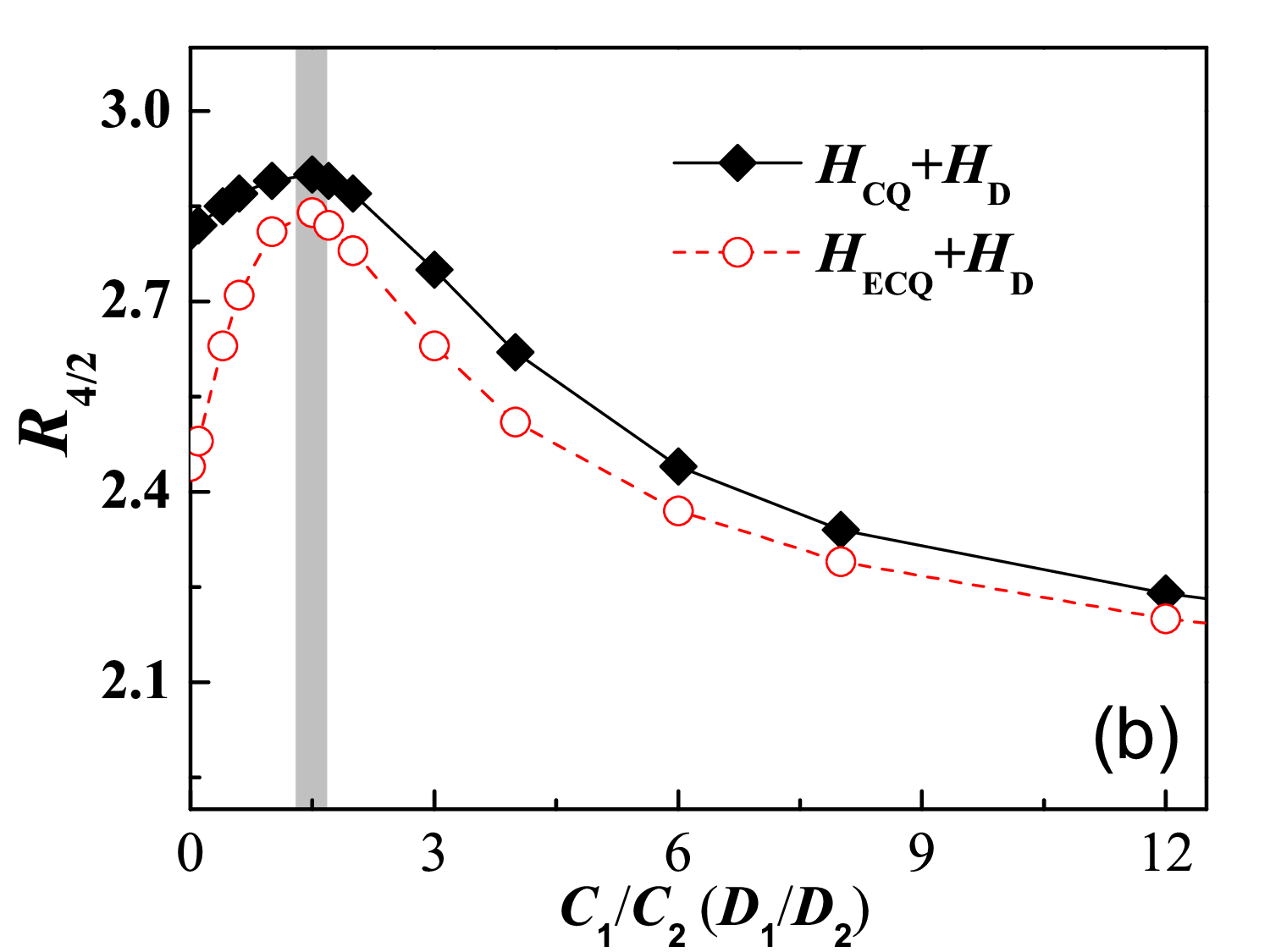}
\caption{(a) The results for $B_{4/2}$, derived from the triaxial models based on the consistent-$Q$ formula ($H_{\mathrm{CQ}}+H_{\mathrm{D}}$) and its extension ($H_{\mathrm{ECQ}}+H_{\mathrm{D}}$), are presented as a function of $D_1/D_2$ or $C_1/C_2$, with all other parameters maintained as in Fig~\ref{F2}. (b) The same as in (a) but for the $R_{4/2}$ ratio.}\label{F22}
\end{center}
\end{figure}

In addition to the mean-field method, one may seek to analyze how the effective $\gamma$ deformation (dynamical triaxiality) influences the $B(E2)$ anomaly feature. For that, we take the consistent-$Q$ Hamiltonian $\hat{H}_\mathrm{S_D}$ and its extension $\hat{H}_\mathrm{S_C}$ as the example to exemplify the derivation of $\hat{H}_\mathrm{D}$ using the second approach described above. In the calculations, the parameters (in any units) are chosen as $(D_1,~D_2,~\chi)=(0.0,~-9.0,~0.23)$ and $(C_1,~C_2,~k3,~\chi)=(0.0,~-9.0,~3.0,~-1.1)$, which in turn yield via Eq.~(\ref{effgamma}) the effective deformation $\gamma_{\mathrm{eff}.}\approx41^\circ$ for both cases. The difference is that triaxial deformation for $\hat{H}_\mathrm{S_C}$ arises from competition between the prolate and oblate shapes as illustrated in Fig.~\ref{F02}(b). With the ground-state wave functions solved from $\hat{H}_\mathrm{S_D}$ and $\hat{H}_\mathrm{S_C}$, we further derive the parameters involved in $\hat{H}_\mathrm{D}$ associated with the inertia parameters, $\Gamma_1:\Gamma_2:\Gamma_3=3:1:4$, and the parameter values for two schemes are obtained as $(a,~b,~c)=(2.66,~0.44,~0.0004)$ and $(a,~b,~c)=(2.65,~0.61,~0.002)$, respectively. To examine the triaxial rotor modes derived from each scheme, the low-lying patterns solved from both $\hat{H}_\mathrm{Tri}=\hat{H}_\mathrm{S}+\hat{H}_\mathrm{D}$ and its static part $\hat{H}_\mathrm{S}$ are presented in Fig.~\ref{F2} to make a comparison. To further investigate the influence of vibration mode on $B(E2)$ anomaly, the $R_{4/2}$ and $B_{4/2}$ ratios are solved from $\hat{H}_\mathrm{Tri}$ with the U(5) term additionally added, and the results are presented as a function of $D_1/D_2$ or $C_1/C_2$, as depicted in Fig.~\ref{F22}.

As observed in Fig.~\ref{F2}(D1), a regular rotational pattern with $B_{4/2}>1.0$ has been generated using the consistent-$Q$ Hamiltonian with the given parameters. Upon incorporating triaxial rotor modes, it is shown that all the energy levels are significantly lowered, particularly those within the $\gamma$ band. This observation suggests a strong band mixing between the ground-state band and the $\gamma$ band. During this variation, the $B(E2)$ ratio shifts from $B_{4/2}=1.41$ to $B_{4/2}=0.65$. As further observed in Fig.~\ref{F2}(D2), the energy levels derived from the extended consistent-$Q$ scheme experience an even more pronounced decrease after adding $\hat{H}_\mathrm{D}$ to the Hamiltonian, which implies that a stronger band mixing will occur in this case. Consequently, notable change can be observed in the ratio values: $R_{4/2}$ decreases from 3.2 to 2.44 and $B_{4/2}$ declines from 1.36 to 0.26, highlighting a more pronounced $B(E2)$ anomaly. By continuously increasing the proportion of the U(5) component in the Hamiltonian, as shown in Fig.~\ref{F22}, one can observe that the $B(E2)$ ratio in both schemes may quickly increase from $B_{4/2}<1.0$ to $B_{4/2}>1.0$, ultimately converging towards the vibrational limit with $B_{4/2}\sim1.7$. In contrast, the energy ratio $R_{4/2}$ as a function of $D_1/D_2$ or $C_1/C_2$ exhibits non-monotonic evolutions before reaching the vibrational limit with $R_{4/2}\sim2.0$. Notably, the maximal values of $R_{4/2}$ appear around $D_1/D_2(C_1/C_2)\approx1.5$ for both cases, which approximately aligns with the "critical point" regarding the $B(E2)$ anomaly where $B_{4/2}\leq1.0$, suggesting a phase transition-like evolution between the normal cases with $B_{4/2}>1.0$ and abnormal cases with $B_{4/2}<1.0$.

Combined with the above mean-field analysis and numerical calculations, the results confirm that $B(E2)$ anomaly can be attributed to be a collective behavior relevant to triaxiality, which is associated with strong band mixing. Such a phenomenon is not expected to be observed in an axially-deformed systems. This point can be partially understood from the SU(3) IBM realization of the rotor modes~\cite{Zhang2014}, where adding $\hat{H}_\mathrm{D}$ does not alter the $B(E2)$ structure linked to the axially symmetric SU(3) IRREPs $(2N, 0)$ or $(0,N)$~\cite{Kotabook}. In addition, one can discern that spectral patterns associated with $B(E2)$ anomaly, as illustrated in Fig.~\ref{F2} and Fig.~\ref{F1}, exhibit a global similarity and consistently feature low-lying yrare bands adjacent to yrast bands. This characteristic may serve as a reference point for developing models that generate $B(E2)$ anomaly within other theoretical frameworks.

\begin{center}
\vskip.2cm\textbf{(c). Application to the neutron-deficient nuclei}
\end{center}\vskip.2cm

\begin{figure}
\begin{center}
\includegraphics[scale=0.33]{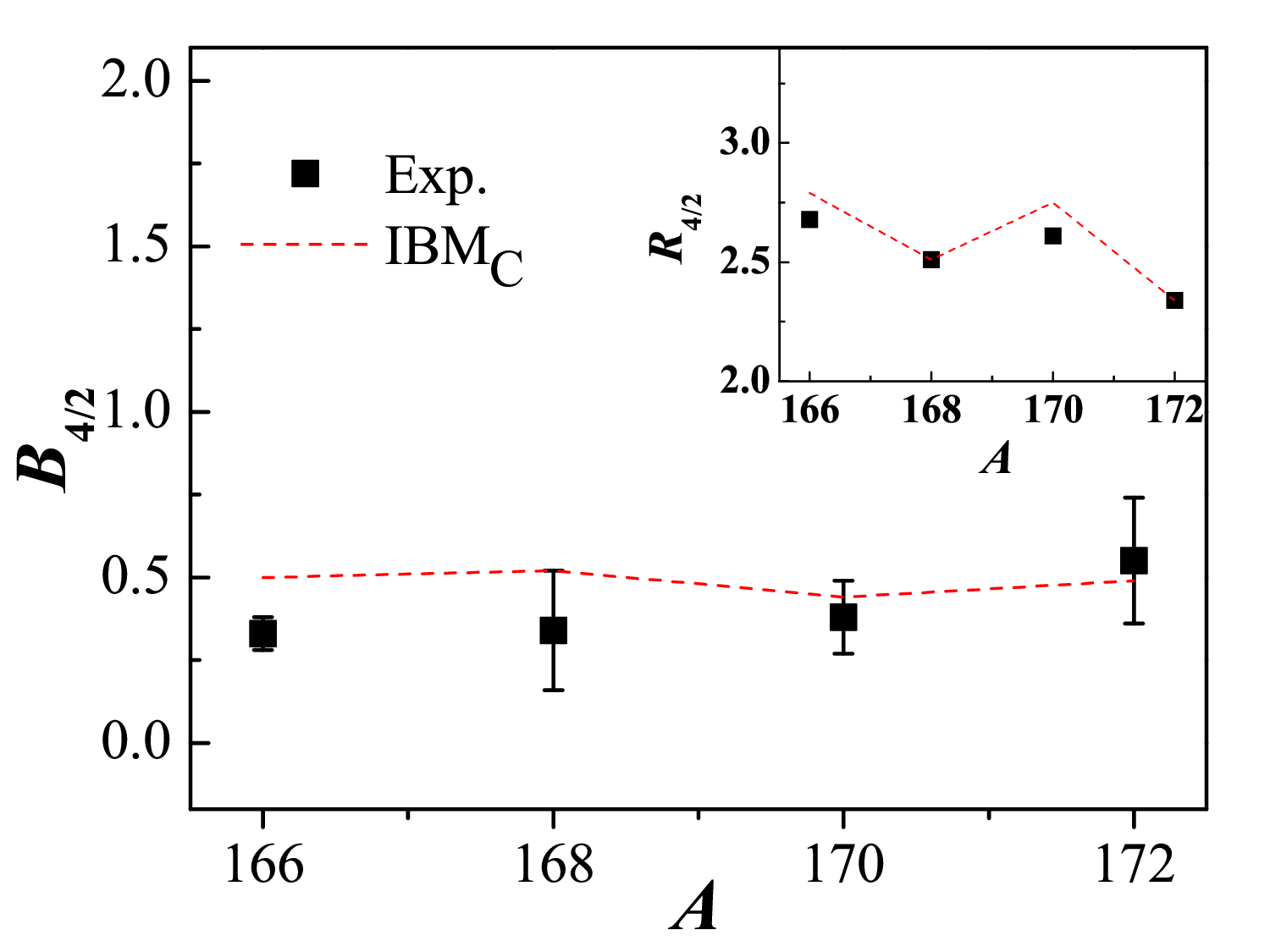}
\caption{The evolution of $R_{4/2}$ and $B_{4/2}$ in $^{166}$W, $^{168,170}$Os and $^{172}$Pt is presented as a function of the mass number $A$ to compare with the model results exacted from Table~\ref{T1}.  }\label{F3}
\end{center}
\end{figure}

To conduct a preliminary examination of the model application, we apply the Hamiltonian constructed on the extended consistent-$Q$ formulism (\ref{HSC}) to describe the available data for the low-lying states in $^{172}$Pt, $^{170}$Os, $^{168}$Os and $^{166}$W. These neutron-deficient nuclei have all been observed to exhibit anomalous $E2$ behaviors in their yrast bands~\cite{Grahn2016,Saygi2017,Cederwall2018,Goasduff2019}. The calculated results and related experimental data are presented in Table~\ref{T1} and Fig.~\ref{F3}. In the calculations, the parameters in $\hat{H}_{\mathrm{S_C}}$ have been set by $C_1/C_2=0.8$ for both $^{172}$Pt and $^{168}$Os corresponding to $N=8$ and by $C_1/C_2=0.6$ for both $^{170}$Os and $^{166}$W corresponding to $N=9$. The dimensionless parameters $k_3$ and $\chi$ are always fixed as $k_3=3.1$ and $\chi=-\sqrt{7}/2$.
This parameterization may yield the effective triaxial deformation $\gamma_{\mathrm{eff}.}\approx40^\circ$. The parameters involved in $\hat{H}_{\mathrm{D}}$ are fully determined by
the mapping scheme based on the second approach described in Sec.II(B) with additional adjustments made to the coefficient in front of $\hat{L}^2$ as well as an overall scale factor to better reproduce the empirical values for $E(2_1^+)$ and other level energies. With these considerations, the parameters (in keV) in the calculations for nuclei with $A=166-172$ are adopted as $(C_1,~C_2,~a,~b,~c)=$ (74.9,~-124.8,~21.7,~5.3,~0.35), (100.2,~-125.2,~31,~10,~0.63), (101.8,~-169.7,~21.7,~6.2,~0.32) and (132.1,~-165.2,~44.5,~15.1,~0.57). In deriving the parameters ($a,~b,~c$) via the mapping formulas (\ref{abc}), the inertial parameters have been set by $\Gamma_1:\Gamma_2:\Gamma_3=5:1:10$, $8:1:17$, $5:1:9$ and $11:1:20$, respectively, which are obtained by reproducing the ratios of the low-lying levels in the experiments using the rotor Hamiltonian (\ref{Hr}). For instance, the ratios $\frac{E(4_1^+)}{E(2_1^+)}=2.40$ and $\frac{E(6_1^+)}{E(2_1^+)}=4.0$ can be directly derived from the rotor Hamiltonian with $\Gamma_1:\Gamma_2:\Gamma_3=11:1:20$. The corresponding experimental values in $^{172}$Pt are reported as $\frac{E(4_1^+)}{E(2_1^+)}=2.34$ and $\frac{E(6_1^+)}{E(2_1^+)}=3.83$.
\begin{table}
\caption{The results (denoted as IBM$_{\mathrm{C}}$) obtained from the model built on the extended consistent-$Q$ formula are presented for comparison with available data of the low-ling level energies (in MeV) and $B(E2)$ transitions (in W.u.) for $^{172}$Pt~\cite{Cederwall2018}, $^{168,170}$Os~\cite{Grahn2016,Goasduff2019}, and $^{166}$W~\cite{Saygi2017}, where "-"
indicates unknown data. The parameters adopted in the calculations are detailed in the text, and the effective charge $e_\mathrm{B}$ is determined by the experimental value of $B(E2;2_1^+\rightarrow0_1^+)$ for each nucleus.}
\begin{center}
\label{T1}
\begin{tabular}{ccc|cc|cc|cc}\hline\hline
Energy&$^{172}$Pt&$\mathrm{IBM_C}$&$^{170}$Os&$\mathrm{IBM_C}$&$^{168}$Os&$\mathrm{IBM_C}$&$^{166}$W&$\mathrm{IBM_C}$\\
\hline
$E(2_1^+)$&0.458&0.458&0.287&0.287&0.341&0.341&0.252&0.252\\
$E(4_1^+)$&1.070&1.071&0.750&0.789&0.857&0.858&0.676&0.703\\
$E(6_1^+)$&1.753&1.793&1.325&1.185&1.499&1.485&1.226&1.135\\
$E(8_1^+)$&2.405&2.401&1.946&1.817&2.223&2.063&1.865&1.782\\
$E(10_1^+)$&-&3.443&2.545&2.195&2.983&2.994&2.552&2.297\\
$E(0_2^+)$&-&0.521&-&0.500&-&0.395&-&0.367\\
$E(2_2^+)$&-&0.611&-&0.405&-&0.460&-&0.341\\
$E(3_1^+)$&-&1.084&-&0.692&-&0.815&-&0.594\\
$E(4_2^+)$&-&1.377&-&0.907&-&1.064&-&0.802\\ \hline
$L_i^+\rightarrow L_f^+$ &$^{172}$Pt&$\mathrm{IBM_C}$&$^{170}$Os&$\mathrm{IBM_C}$&$^{168}$Os&$\mathrm{IBM_C}$&$^{166}$W&$\mathrm{IBM_C}$\\ \hline
$2_1^+\rightarrow0_1^+$&49($11$)&49&97$_{-9}^{+9}$&97&74(13)&74&150(9)&150\\
$4_1^+\rightarrow2_1^+$&27($7$)&24&38$_{-7}^{+13}$&43&25(13)&39&50(7)&75\\
$6_1^+\rightarrow4_1^+$&-&20&-&39&-&33&18(4)&65\\
$8_1^+\rightarrow6_1^+$&-&36&-&11&-&52&-&19\\
$0_2^+\rightarrow2_1^+$&-&28&-&7.0&-&27&-&15\\
$2_2^+\rightarrow0_1^+$&-&9.8&-&0.5&-&6.6&-&2.3\\
$2_2^+\rightarrow2_1^+$&-&82&-&116&-&124&-&199\\
$3_1^+\rightarrow2_2^+$&-&74&-&162&-&110&-&248\\
$4_1^+\rightarrow2_2^+$&-&8.1&-&22&-&16&-&31\\
$4_2^+\rightarrow2_1^+$&-&11&-&78&-&46&-&115\\
$4_2^+\rightarrow3_1^+$&-&10&-&7.4&-&58&-&117\\
\hline\hline
\end{tabular}
\end{center}
\end{table}

As illustrated in Table~\ref{T1}, the available data for these neutron-deficient nuclei can generally be well described by the model calculations. These calculations additionally predict low-energy non-yrast states, such as $0_2^+$, $2_2^+$ and $4_2^+$, indicating triaxiality cuased by the large effective $\gamma$ deformation. The assumption of  triaxial deformation is consistent with the previous mean-field analysis of the relevant nuclei \cite{Guzman2010}. More importantly, the significantly suppressed $B_{4/2}$ ratio as observed in all four nuclei can be reasonably explained from the model results, which meanwhile suggest that the anomalous $E2$ behavior may persist even for yrast states with $L>4$ like the result identified from $^{166}$W. Moreover, it is shown that $B(E2;6_1^+\rightarrow4_1^+)=18(4)$ W.u. in $^{166}$W is overestimated by the calculated result (65 W.u.), partially because of a large effective charge adopted in this case to reproduce the experimental $B(E2;2_1^+\rightarrow0_1^+)$ value, which is notably higher than those of observed in other cases. To some extent, this qualitative deviation can be mitigated by relaxing the rotor mapping constraints on the parameters in describing experiments. Based on the aforementioned theoretical analysis~\cite{Zhang2022,Zhang2014}, the suppressed $B(E2)$ cascade for the yrast bands can be attributed to band mixing effects, which align with the predicted strong inter-band $E2$ transitions between members of yrast band and those of yrare band, as shown in Table~\ref{T1}. For instance, the calculated results for $^{166}$W even predict another anomaly feature where $B(E2;4_2^+\rightarrow2_1^+)>B(E2;4_1^+\rightarrow2_1^+)$, a phenomenon that has not been observed in conventional modes. All these predictions actually suggest a robust test of the present theoretical analysis in future experiments. The consistency between the experiments and theoretical calculations is further evidenced by the evolution of $B_{4/2}$ and $R_{4/2}$, as depicted in Fig.~\ref{F3}. In addition to the $B(E2)$ anomaly characterized by $B_{4/2}<1.0$, it is shown that a slight fluctuation in $R_{4/2}$ alongside a nearly constant change in $B_{4/2}$ with increasing mass number $A$ has been reproduced rather well in theory.

The above descriptions provide further substantiation for the previously proposed viewpoint~\cite{Zhang2022} that the $B(E2)$ anomaly in the experiments could be simply elucidated through triaxial rotor modes. In a separate study~\cite{Wang2020}, a Hamiltonian similar to the extended consistent-$Q$ formula employed herein was utilized to fit low-lying data in $^{170}$Os, where the observed depressed value of $B_{4/2}$ was interpreted as resulting from the prohibition of $E2$ transitions between two distinct SU(3) IRREPs. In contrast, our current analysis, based on a more comprehensive approach for deriving rotor modes, demonstrates that the $B(E2)$ anomaly in the IBM is fundamentally a triaxiality-related phenomenon without being necessarily limited to a specific symmetry or model Hamiltonian.
Although the results for $^{170}$Os presented in Table~\ref{T1} are generally consistent with those obtained in \cite{Wang2020}, which employed freely adjustable parameters, we note an overestimation of $E(10_1^+)\approx3.1$ MeV in comparison with the experimental data $E(10_1^+)\approx2.5$ MeV and the present calculation yielding $E(10_1^+)\approx2.2$ MeV. Another recent study~\cite{Pan2024} utilized the consistent-$Q$ formula augmented with high-order terms, similar to that used in Fig.~\ref{F2}(d1), but with freely adjustable parameters to fit data for more neutron-deficient nuclei. This approach results in some conclusions being drawn in a parameter-dependent manner, similar to other previous studies~\cite{Zhang2022,Wang2023,Wang2020} that rely on specific Hamiltonian. The primary objective of the current study, as mentioned at the outset, is to provide a comprehensive understanding of $B(E2)$ anomaly within the IBM. Clearly, the present analysis, based on more general triaxial schemes and rotor mapping procedure without restricting to any particular types of Hamiltonian, offers a more transparent geometric explanation of $B(E2)$ anomaly, thereby clarifying confusing aspects of modeling this exotic phenomenon within the IBM framework.

Generally speaking, the results obtained from different models concerning a given nucleus associated with $B(E2)$ anomaly are qualitatively similar. For instance, all the models~\cite{Zhang2022,Wang2020} predict the low-energy non-yrast states and their generally strong $E2$ connections to yrast states, which have not yet observed in relevant nuclei. Some quantitative differences can be mainly attributed to parameter elaboration across different models. For example, the current calculations for $^{168}$Os predict a lower excitation energy of the $0_2^+$ state at $E(0_2^+)\approx0.395$ MeV and thus a stronger $E2$ transition to the yrast state with $B(E2;0_2^+\rightarrow2_1^+)\approx27$ (in W.u.). This is contrast to the corresponding results (denoted by IBM$_b$) presented in Table 1 of Ref.~\cite{Zhang2022}, which report $E(0_2^+)\approx0.568$ MeV and $B(E2;0_2^+\rightarrow2_1^+)\approx0.3$ (in W.u.). Conversely, results denoted by IBM$_a$, also listed in Table 1 of Ref.~\cite{Zhang2022}, suggest that adopting more stringent constraints on parameters in the SU(3) model yields values of $E(0_2^+)\approx0.287$ MeV and $B(E2;0_2^+\rightarrow2_1^+)\approx18$ (in W.u.), which are closer to the present results. A similar situation is also observed in calculations for $^{172}$Pt. Notably, a constraint of parameter from rotor mapping will make it easier to get a regular yrast level structure consistent with the experimental situation. For example, the results in Table~\ref{T1} indicate that
the typical energy ratios $\frac{E(4_1^+)}{E(2_1^+)}$ and $\frac{E(6_1^+)}{E(2_1^+)}$ for $^{168}$Os are given by $2.51$ and $4.40$, respectively, consisting well with
the present calculations, $2.52$ and $4.35$. While, the corresponding theoretical results obtained in \cite{Pan2024} with the model parameters determined from a global fit for data are suggested to be $\frac{E(4_1^+)}{E(2_1^+)}=2.03$ and $\frac{E(6_1^+)}{E(2_1^+)}=2.77$ for this nucleus. All these differences may provide additional references for the future measurements involving relevant nuclei.

In addition, collective modes exhibiting $B_{4/2}<1.0$ have been also identified in intermediate-mass nuclei~\cite{Kintish2014} and even light nuclei~\cite{Tobin2014}. The associated triaxial deformations are shown to be in good agreement with the no-core symplectic shell model calculations~\cite{Tobin2014} as well the related mean-field analysis \cite{Mitra2002}. Another point to emphasize is that, although $B(E2)$ anomaly could be regarded as an indicator of triaxiality in even-even nuclei based on the current analysis, triaxial deformation does not necessarily imply the presence of $B(E2)$ anomaly. Typical examples include $^{190,192}$Os, where the properties of the low-lying states along with the normal ratio $B_{4/2}>1.0$ can also be well recognized from the triaxial rotor modes~\cite{Allmond2008,Zhang2024}. The primary distinction lies in the fact that significantly stronger band-mixing effects are anticipated to occur in nuclei associated with $B(E2)$ anomaly. $^{76}$Ge is another noteworthy instance proposed to exhibit triaxial deformation~\cite{Toh2013,Ayangeakaa2023} irrelevant to $B(E2)$ anomaly. It was suggested that $^{76}$Ge might serve as a rare example of an even-even nucleus showing rigid triaxial deformation, evidenced by the observed appropriate level staggering within its $\gamma$ band~\cite{Toh2013}. Potentially,  $^{76}$Ge may serve as an experimental candidate for validating the algebraic methodologies proposed in this work for deriving triaxial rotor model, which will be discussed elsewhere.

\begin{center}
\vskip.2cm\textbf{V. Summary}
\end{center}\vskip.2cm

In summary, a general procedure for constructing triaxial rotor modes in the IBM has been demonstrated without imposing any symmetries. Based on this procedure, an thorough analysis of the triaxial modes derived from different schemes is conducted to elucidate the collective mechanism underlying the $B(E2)$ anomaly phenomena. It is shown that the anomalous $E2$ behavior with $B_{4/2}<1.0$ can manifest in all the triaxial (whether intrinsic or dynamic) schemes and is thus posited as a robust triaxility-relevant phenomenon in the IBM. This perspective is further examined through the application of the model constructed on the extended consistent-$Q$ formula to describe some neutron-deficient nuclei associated with $B_{4/2}<1.0$. The results support a simple rotor mode interpretation of the observed $B(E2)$ anomaly phenomenon, thereby broadening the IBM descriptions to the exotic nuclei with extreme ratio between proton number and neutron number. An important prediction is that there should exist many low-energy non-yrast states in the $B(E2)$ anomaly systems, which provides a critical criterion for evaluating the current theory in future experimental investigations. However, it remains imperative to explore why such triaxial modes appear only in these neutron-deficient nuclei through more microscopic approaches. In addition to the even-even neutron-deficient nuclei discussed above, similar observations of the $B(E2)$ anomaly phenomenon have also been reported~\cite{Zhang2021} in adjacent odd-A species. How to extend the present analysis to odd-A systems and to the microscopic version of the IBM~\cite{Iachellobook,Caprio2004,Dieperink1982}, particularly the microscopic mean-field derivation of the IBM Hamiltonian~\cite{Nomura2008,Nomura2010,Vasileiou2024} would be interesting and highly expected. Related work is in progress.
\bigskip

\begin{acknowledgments}
Fruitful discussions with F. Pan, C. Qi, T. Wang, J. P. Draayer, and L. Fortunato are gratefully acknowledged.
Support from the Natural Science Foundation of China (12375113,11875158) is acknowledged.
\end{acknowledgments}



\end{document}